\documentclass[12pt]{article}

\usepackage[toc,page]{appendix}
\usepackage{slashed}
\usepackage[sumlimits,intlimits,namelimits]{amsmath} 
\usepackage[english]{babel}
\usepackage{graphicx}
\usepackage{hyperref}
\usepackage{cite}

\newcommand{\ba}{\begin{eqnarray}}
\newcommand{\ea}{\end{eqnarray}}

\newcommand{\ice}[1]{\relax}

\usepackage{color}
\usepackage{amsmath}
\usepackage{booktabs} 
\usepackage{subfigure}

\DeclareMathOperator{\Tr}{Tr}

\def\Li{\mbox{Li}}

\begin{document}

\begin{titlepage}

\begin{flushright}
SI-HEP-2022-19\\
SFB-257-P3H-22-086
\end{flushright}
\vspace{1.2cm}
\begin{center}
	  { \Large\bf NLO QCD corrections to inclusive semitauonic weak decays of heavy hadrons up to $1/m_b^3$  }
\end{center}
\vspace{0.5cm}
\begin{center}
{\sc Daniel Moreno} \\[0.2cm]
{\sf Center for Particle Physics Siegen, Theoretische Physik 1, Universit\"at Siegen\\ 57068 Siegen, Germany}
\end{center}

\vspace{0.8cm}
\begin{abstract}\noindent
 This paper presents $\alpha_s$ corrections to the inclusive semitauonic $B \to X_c \tau \bar{\nu}_\tau$ 
 decay width and leptonic invariant mass spectrum up to order $1/m_b^3$ in 
 the heavy quark expansion. Analytical results with full dependence on the final-state quark and tau lepton masses are obtained.  
 The results up to $1/m_b^2$ can be extrapolated to the $B \to X_u \tau \bar{\nu}_\tau$ decay by taking the limit of 
 vanishing final-state quark mass.
\end{abstract}

\end{titlepage}

\section{Introduction} 
\label{sec:Intro}

The study of inclusive weak decays of heavy flavoured hadrons at the precision level 
is important for testing the flavour sector of the Standard Model (SM). 
On the one hand, they allow for a precise extraction of Cabbibo-Kobayashi-Maskawa (CKM) matrix elements, 
which in turn is required to understand current tensions between data and the SM predictions ($B$-anomalies), which could 
hint the presence of new physics. On the other hand, describing the pattern of lifetimes and lifetime differences constitute 
a solid test of our understanding of the strong interactions in its interplay with the weak interactions.

In particular, one of the $B$-anomalies is the current experimental values for the ratios 
$R(D^{(*)}) = \mathcal{B}(B\rightarrow D^{(*)}\tau \bar{\nu}_\tau)/\mathcal{B}( B\rightarrow D^{(*)} e \bar{\nu}_e)$ in 
$B$ decays mediated by the $b\rightarrow c \ell \bar{\nu}_\ell$ transition ($\ell =e,\mu,\tau$), which show a more than $3\sigma$ deviation from the 
standard model prediction\cite{BaBar:2012obs,BaBar:2013mob,LHCb:2015gmp,Belle:2015qfa,Belle:2017ilt,LHCb:2017rln,Belle:2019rba,HFLAV:2022pwe,Bernlochner:2021vlv}. 
This fact might point out the presence of new physics coupled to the $\tau$ lepton, 
which is not present in its relatives $e$ and $\mu$, due to its large mass. As 
suggested in~\cite{Ligeti:2014kia,Ligeti:2021six}, the inclusive decays can provide valuable 
complementary information to the one of exclusive decays, which motivates also their investigation. 
The inclusive semitauonic decays have been poorly measured so far, 
as they are challenging decays for the experiment.
However it should be possible to measure them in the near future by Belle II.

Inclusive weak decays of heavy flavoured hadrons can be computed within an operator product expansion by 
taking advantage of the fact that $m_b\gg\Lambda_{\rm QCD}$, so that the heavy quark contained in the heavy hadron is almost on shell. 
Therefore, the use of heavy quark effective theory (HQET)~\cite{Shifman:1987rj,Eichten:1989zv,Isgur:1989vq,Grinstein:1990mj} to separate hard 
modes (of order $m_b$) and soft modes (of order $\Lambda_{\rm QCD}$) in the hadronic system is natural. 
The approach used for the separation of perturbative and non-perturbative 
effects in the expression for the decay rate and kinematical distributions is the so-called heavy quark 
expansion (HQE)~\cite{Chay:1990da,Bigi:1992su,Bigi:1993fe,Blok:1993va,Manohar:1993qn}. 
The result is an expansion in $\Lambda_{\rm QCD}/m_b$ and $\alpha_s(m_b)$ 
which allows for a systematic improvement of theoretical predictions by pushing for higher orders in the expansion 
parameters.

The HQE for the inclusive semitauonic decay rate and distributions is already quite developed for both, the 
massive and massless final state quark cases. 
The leading term is known at next-to-next-to-leading order (NNLO)-QCD~\cite{Ho-kim:1983klw,Czarnecki:1994bn,Jezabek:1996db,Jezabek:1997rk,Biswas:2009rb}. 
The leading $(\Lambda_{\rm QCD}/m_b)^2$ non-perturbative corrections are 
known at leading order (LO)-QCD~\cite{Balk:1993sz,Koyrakh:1993pq,Falk:1994gw,Ligeti:2014kia,Ligeti:2021six}. 
Finally, the next-to-leading $(\Lambda_{\rm QCD}/m_b)^3$ non-perturbative corrections have been considered to 
LO-QCD for the case of massive final state quark~\cite{Mannel:2017jfk,Colangelo:2020vhu,Rahimi:2022vlv}.

This paper addresses the computation of the $(\Lambda_{\rm QCD}/m_b)^2$ and $(\Lambda_{\rm QCD}/m_b)^3$ non-perturbative corrections 
to NLO-QCD for the total rate and the leptonic invariant mass spectrum in the $B\rightarrow X_c \tau \bar{\nu}_\tau$ decay, analytically. 
The $(\Lambda_{\rm QCD}/m_b)^2$ results can be extrapolated to the $B\rightarrow X_u \tau \bar{\nu}_\tau$ decay 
by taking the limit of vanishing final state quark mass. Moments of the spectrum can be readily obtained by integrating 
the differential rate with the desired weight function. 
This paper is an extension of~\cite{Mannel:2021zzr} where the same quantities were computed for the case of massless 
leptons $B\rightarrow X_c e\bar{\nu}_e$.

A Mathematica notebook called ``cobqtv.nb'' is provided as an ancillary file, which contains analytical results for the 
coefficients of the total rate and the leptonic invariant mass spectrum up to order $\alpha_s/m_b^3$ 
for $B\rightarrow X_c \tau \bar{\nu}_\tau$ and up to order $\alpha_s/m_b^2$ for $B\rightarrow X_u \tau \bar{\nu}_\tau$.
The file can be downloaded from arXiv by using the link in ``ancillary files'' or by downloading the entire source package as a 
gzipped tar file (.tar.gz).

The paper is organized as follows. Sec.~\ref{sec:HQEIDHH} gives the definitions for the HQE of the total rate~\ref{subsec:rate} and 
the leptonic invariant mass spectrum~\ref{subsec:difrate}. 
Sec.~\ref{sec:HQE} describes the outline of the calculation and presents the $\alpha_s(m_b)$ corrections to the partonic~\ref{difratemb0}, 
$(\Lambda_{\rm QCD}/m_b)^2$~\ref{difratemb2}, and $(\Lambda_{\rm QCD}/m_b)^3$~\ref{difratemb3} 
coefficients in the HQE of the decay spectrum and total rate. 
Finally, sec.~\ref{sec:disc} is devoted to a numerical analysis to discuss the impact of the new results.

\section{HQE for inclusive decays of heavy flavoured hadrons}
\label{sec:HQEIDHH}

\subsection{The total rate}
\label{subsec:rate}

This section briefly describes the theoretical framework used for the calculation of inclusive semileptonic decays of heavy hadrons 
and provides the main definitions. At low momentum transfer compared to $M_W$ the heavy quark decay $b\rightarrow q \tau \bar\nu_\tau$ can be described 
by an effective Fermi Lagrangian 
\begin{equation}
  \label{eq:FermiLagr}
{\cal L}_{\rm eff} = 2\sqrt{2}G_F \sum_{q= c,u} V_{qb}(\bar{b}_L \gamma_\mu q_L) 
(\bar{\nu}_{\tau, L} \gamma^\mu \tau_L) + {\rm h.c.} \, , 
\end{equation}
where the subscript $L$ denotes left-handed fermionic fields, $G_F$ is the Fermi constant 
and $V_{qb}$ is the corresponding CKM matrix element. 

By using the optical theorem, the inclusive decay rate of
$B\to X_q\tau \bar{\nu}_\tau$ is obtained from the imaginary part of the forward hadronic matrix element of the transition operator 
${\cal T}$
\begin{equation}\label{eq:trans_operator}
{\cal T} = i \int d^4 x\,    
T\left\{ {\cal L}_{\rm eff} (x),  {\cal L}_{\rm eff} (0) \right\} \, ,
\quad \Gamma (B \to X_q \tau \bar{\nu}_\tau)
= \frac{1}{2M_B} \text{Im }\langle B|{\cal T} |B\rangle \,,
\end{equation} 
where $M_B$ is the heavy hadron mass. Since the heavy quark mass $m_b$ (hard scale) is much larger than the hadronization 
scale of QCD $\Lambda_{\rm QCD}$ (soft scale) the equation above contains perturbatively calculable 
contributions which can be factorized from the non-perturbative ones by using an operator product expansion, the so-called 
HQE (see e.~g.~\cite{Mannel:2014xza,Mannel:2015wsa,Mannel:2015jka,Mannel:2021zzr} for more details). Then the $\mbox{Im}\, {\cal T}$, computed as an expansion near the heavy 
quark mass shell in QCD, is matched to an expansion in $\Lambda_{\rm QCD}/m_b$ by using local operators in HQET~\cite{Mannel:1991mc,Manohar:1997qy}.
\begin{eqnarray}
	\label{eq:HQE-1}
	\mbox{Im}\, \mathcal{T} = \Gamma^0 |V_{qb}|^2 
	\bigg( C_0 \mathcal{O}_0 
	+ C_v \frac{\mathcal{O}_v}{m_b} 
	+ C_\pi \frac{\mathcal{O}_\pi}{2m_b^2} 
	+ C_G \frac{\mathcal{O}_G}{2m_b^2} 
	+ C_D \frac{\mathcal{O}_D}{4m_b^3}
	+ C_{LS} \frac{\mathcal{O}_{LS}}{4m_b^3}
	\bigg)\,,
\end{eqnarray}
where $\Gamma^0 = G_F^2 m_b^5/(192 \pi^3)$, $C_i$ are the matching coefficients which have a perturbative expansion in the 
strong coupling constant $\alpha_s (m_b)$, and $\mathcal{O}_i$ are the HQET operators listed below
\begin{eqnarray}
 \mathcal{O}_0 &=& \bar h_v h_v \qquad\qquad\qquad\qquad\qquad\qquad\;\;\, \mbox{(Leading power operator)}\,,
 \\
 \mathcal{O}_v &=& \bar h_v v\cdot \pi h_v \qquad\qquad\qquad\qquad\qquad\;\;\; \mbox{(EOM operator)}\,,
 \\
 \mathcal{O}_\pi &=& \bar h_v \pi_\perp^2 h_v \qquad\qquad\qquad\qquad\qquad\quad\;\; \mbox{(Kinetic operator)}\,,
 \label{mupi} \\
 \mathcal{O}_G &=& \frac{1}{2}\bar h_v [\gamma^\mu, \gamma^\nu] \pi_{\perp\,\mu}\pi_{\perp\,\nu}  h_v \quad\quad\qquad\quad\;\; \mbox{(Chromomagnetic operator) }\,,
 \label{muG} \\
 \mathcal{O}_D &=& \bar h_v[\pi_{\perp\,\mu},[\pi_{\perp}^\mu , v\cdot \pi]] h_v \qquad\qquad\qquad \mbox{(Darwin operator)}\,,
 \label{rhoD}\\
 \mathcal{O}_{LS} &=& \frac{1}{2}\bar h_v[\gamma^\mu,\gamma^\nu]\{ \pi_{\perp\,\mu},[\pi_{\perp\,\nu}, v\cdot \pi] \} h_v \quad\, \mbox{(Spin-orbit operator)}\,. 
 \label{Ops} \label{rhoLS}
\end{eqnarray}
Here $\pi_\mu = i D_\mu = i\partial_\mu +g_s A_\mu^a T^a$ is the QCD covariant derivative, 
$a^\mu_\perp = a^\mu - v^\mu (v\cdot a)$ with $v=p_B/M_B$ being the heavy hadron velocity, 
and $h_v$ is the HQET field whose dynamics is determined by the HQET Lagrangian~\cite{Manohar:1997qy}. 
Operators which are of higher dimension on shell have been neglected.

It is convenient to trade the leading term operator $\mathcal{O}_0$ in Eq.~(\ref{eq:HQE-1}) by the local QCD operator 
$\bar b \slashed v b$, since its forward hadronic matrix element is completely normalized. To that purpose, 
the HQE of the $\bar b \slashed v b$ operator is needed up to the desired order
\begin{equation}
\bar b \slashed v b = \mathcal{O}_0 + \tilde{C}_v \frac{\mathcal{O}_v}{m_b} + \tilde C_\pi \frac{\mathcal{O}_\pi}{2m_b^2} + \tilde C_G \frac{\mathcal{O}_G}{2m_b^2} 
 + \tilde C_D \frac{\mathcal{O}_D}{4m_b^3}
 + \tilde C_{LS} \frac{\mathcal{O}_{LS} }{4m_b^3} \,,
 \label{hqebvb}
\end{equation}
where $\tilde{C}_i$ are the matching coefficients which are pure numbers.

In addition, the equation of motion (EOM) of the HQET 
Lagrangian is used for the operator $\mathcal{O}_v$ in Eq.~(\ref{eq:HQE-1}).

\begin{eqnarray}
  \mathcal{O}_v =
 - \frac{1}{2m_b} (\mathcal{O}_\pi+ c_F(\mu)\mathcal{O}_G)
 -  \frac{1}{8m_b^2} (c_D(\mu)\mathcal{O}_D + c_S(\mu)\mathcal{O}_{LS})\,,
 \label{LHQET}
\end{eqnarray}
where

\begin{eqnarray}
 c_F(\mu) &=& 1 + \frac{\alpha_s}{2\pi}\bigg[ C_F + C_A\bigg(1 + \ln\left(\frac{\mu}{m_b}\right)\bigg) \bigg]\,,
 \nonumber
 \\
 c_D(\mu) &=& 1 + \frac{\alpha_s}{\pi}\bigg[ 
 C_F\bigg(-\frac{8}{3}\ln\left(\frac{\mu}{m_b}\right) \bigg) 
 + C_A\bigg(\frac{1}{2} - \frac{2}{3}\ln\left(\frac{\mu}{m_b}\right)\bigg) \bigg]\,,
\end{eqnarray}
are the coefficients of the chromomagnetic and Darwin operators in the HQET Lagrangian at NLO~\cite{Manohar:1997qy}. Note that, 
due to reparametrization invariance~\cite{Luke:1992cs}, the coefficient of the spin-orbit operator is linked to the one of the 
chromomagnetic operator $c_S = 2c_F -1$. The parameter $\mu$ is the renormalization scale and $C_F=4/3$, $C_A=3$ are colour factors.

Finally, the HQE for inclusive semitauonic weak decays is written as
\begin{eqnarray}
\label{hqewidth2}
\Gamma(B\rightarrow X_q \tau \bar \nu_\tau )
  &=& \Gamma^0 |V_{qb}|^2 
 \bigg[ C_0 \bigg( 1 
- \frac{\bar{C}_\pi - \bar{C}_v }{C_0}\frac{\mu_\pi^2}{2m_b^2}\bigg)
+ \bigg(\frac{\bar{C}_G}{c_F(\mu)} -  \bar{C}_v \bigg)\frac{\mu_G^2}{2m_b^2}
\nonumber
\\
&&
- \bigg(\frac{\bar{C}_D}{c_D(\mu)}-\frac{1}{2} \bar{C}_v \bigg) \frac{\rho_D^3}{2m_b^3}
 - \bigg(\frac{\bar{C}_{LS} }{c_S(\mu)} - \frac{1}{2} \bar{C}_v \bigg) \frac{\rho_{LS}^3}{2m_b^3}
 \bigg]
 \\
 &\equiv& \Gamma^0 |V_{qb}|^2 
 \bigg(   C_0 
- C_{\mu_\pi}\frac{\mu_\pi^2}{2m_b^2}
+ C_{\mu_G}\frac{\mu_G^2}{2m_b^2}
- C_{\rho_D} \frac{\rho_D^3}{2m_b^3}
 - C_{\rho_{LS}} \frac{\rho_{LS}^3}{2m_b^3}
 \bigg)
  \, , 
  \label{hqewidth}
\end{eqnarray}
where $\bar{C}_i\equiv C_i - C_0 \tilde{C}_i$ are defined as the difference 
between the coefficients $C_i$ of the HQE of the transition operator in Eq.~(\ref{eq:HQE-1}) and the current 
in Eq.~(\ref{hqebvb}) multiplied by $C_0$. In the $b \rightarrow c \tau \bar \nu_\tau$ transition, 
the matching coefficients $C_i$ ($i=0,\,\mu_\pi,\,\mu_G,\,\rho_D,\,\rho_{LS}$) depend on two ratios $\rho= m_c^2/m_b^2$ and $\eta = m_\tau^2/m_b^2$, where $m_c$ is the charm quark mass and 
$m_\tau$ is the tau lepton mass. In the $b \rightarrow u \tau \bar \nu_\tau$ transition the matching coefficients only depend on $\eta$ 
since the up quark mass is neglected. 
Note that reparametrization invariance also links the coefficients of the HQE of the rate, 
$C_0 = C_{\mu_\pi}$ and $C_{\mu_G}=C_{\rho_{LS}}$~\cite{Manohar:2010sf,Becher:2007tk,Mannel:2018mqv}.
The HQE hadronic parameters $\mu_\pi^2$, $\mu_G^2$, $\rho_D^3$ and $\rho_{LS}^3$ are defined as the following forward matrix elements 
of HQET local operators taken between full QCD states~\cite{Mannel:2018mqv}.
\begin{eqnarray}
 \langle B(p_B)\lvert \bar b \slashed v b \lvert B(p_B)\rangle &=& 2M_B\,,  \\
 - \langle B(p_B)\lvert \mathcal{O}_\pi \lvert B(p_B)\rangle &=& 2M_B \mu_\pi^2\,, \\ 
 c_F(\mu)\langle B(p_B)\lvert \mathcal{O}_G \lvert B(p_B)\rangle
 &=& 2M_B \mu_G^2\,, \\
  - c_D(\mu)\langle B(p_B)\lvert \mathcal{O}_D \lvert B(p_B)\rangle&=& 4M_B \rho_D^3\,, \\
 -  c_S(\mu)\langle B(p_B)\lvert \mathcal{O}_{LS} \lvert B(p_B)\rangle&=& 4 M_B \rho_{LS}^3\,.
\end{eqnarray}

\subsection{The spectrum on the dilepton invariant mass $q^2$}
\label{subsec:difrate}

This section briefly describes the construction of the HQE for the spectrum on the dilepton invariant mass $q^2$ starting from the expression for
total rate Eq.~(\ref{eq:trans_operator}). The approach described in~\cite{Mannel:2021ubk} is followed, which takes advantage of the fact that 
the leptonic tensor is unaffected by QCD corrections, so it always appears factorized from the hadronic tensor. Therefore, 
one can use a dispersion representation defined in dimensional regularization~\cite{Groote:1999zp} to write the leptonic tensor as an integral differential in $q^2$. 
For the case of a massive lepton and a massless antineutrino the leptonic tensor reads

\begin{eqnarray}
\label{spectrum1}
 &&i\int\frac{d^D k}{(2\pi)^D}\frac{-\Tr(\Gamma^\sigma i(\slashed k + \slashed \ell + m_\tau)\Gamma^\rho i\slashed k)}{k^2((k+\ell)^2 - m_\tau^2)}  
 \\
  &&\quad\quad =  
 \int_{m_\tau^2}^{\infty} d(q^2) \frac{1}{q^2 - \ell^2-i\eta} 
 \frac{1}{(4\pi)^{D/2}}\frac{\Gamma(D/2 - 1)}{\Gamma(D - 2)} \frac{D-2}{D-1}
 \nonumber
 \\
 &&
 \quad\quad\quad \times
 (q^2)^{D/2 - 2} \bigg(1- \frac{m_\tau^2}{q^2}\bigg)^{D - 2} 
   \bigg[
   \bigg( 1 + \frac{D}{D-2}\frac{ m_\tau^2 }{q^2} \bigg) \ell^\rho \ell^\sigma
  - \bigg(  1 + \frac{1}{D-2}\frac{m_\tau^2}{q^2} \bigg) \ell^2 g^{\rho\sigma} 
  \bigg]\,,
  \nonumber
\end{eqnarray}
where $D = 4 -2 \epsilon$ is the number of spacetime dimensions, $\Gamma_\mu  = \gamma_\mu \frac{1}{2}(1-\gamma_5)$, and $\ell$ is the 
four-momentum flowing through the leptons.
Note that, as long as the dispersion representation is defined in dimensional regularization, there is no need for subtractions since singularities 
are properly regularized. Indeed only the imaginary part of Eq.~(\ref{spectrum1}) is required in the present calculation. Since taking the imaginary 
part makes the integral non-singular, commutation of the integral over $q^2$ with expansion in $\epsilon$ and imaginary part 
is justified and it can be conveniently taken

\begin{eqnarray}
 &&\mbox{Im } i\int\frac{d^D k}{(2\pi)^D}\frac{-\Tr(\Gamma^\sigma i(\slashed k + \slashed \ell + m_\tau)\Gamma^\rho i\slashed k)}{k^2((k+\ell)^2 - m_\tau^2)}  
    \label{spectrep}
  \\
  &&\quad\quad = \int_{m_\tau^2}^{\infty}d(q^2) \mbox{Im } \frac{1}{q^2 - \ell^2 -i\eta}
  \frac{1}{24\pi^2}\bigg(1- \frac{m_\tau^2}{q^2}\bigg)^2
  \bigg[
   \bigg(1 + \frac{2m_\tau^2}{q^2}\bigg)  \ell^\rho \ell^\sigma
  -  \bigg(1 + \frac{m_\tau^2}{2q^2} \bigg) \ell^2 g^{\rho\sigma} \bigg] + \mathcal{O}(\epsilon)\,.
  \nonumber
\end{eqnarray}
Let us stress that, whereas this is technically not necessary, it is useful as it simplifies the calculation. 
Note that it is enough to expand the right hand side of Eq.~(\ref{spectrep}) to $\mathcal{O}(\epsilon^0)$ 
since renormalization can be performed for the differential rate, as it holds at the level of the hadronic tensor.

Also note that, after using the dispersion representation, the leptonic tensor becomes 
an ``effective massive propagator'' of mass $q$ and the dependence on $m_\tau$ factorizes from the hadronic part. 
Therefore, the master integrals necessary for the computation of the differential rate are the same that appear in 
the $b \rightarrow c e \bar{\nu}_e$ transition, where $e$ is considered to be massless. These master integrals were computed analytically 
in \cite{Mannel:2021ubk} and they are used for the calculation of the $q^2$ spectrum. 
The HQE for the dilepton invariant mass spectrum can be written in analogy to the HQE for the total rate

\begin{eqnarray}
 \label{hqedifwidth}
\frac{d\Gamma(B\rightarrow X_q \tau \bar{\nu}_\tau )}{dr}
  &=& \Gamma^0 |V_{qb}|^2 
 \bigg[ \mathcal{C}_0 \bigg( 1 
- \frac{\bar{\mathcal{C}}_\pi - \bar{\mathcal{C}}_v }{\mathcal{C}_0}\frac{\mu_\pi^2}{2m_b^2}\bigg)
+ \bigg(\frac{\bar{\mathcal{C}}_G}{c_F(\mu)} -  \bar{\mathcal{C}}_v \bigg)\frac{\mu_G^2}{2m_b^2}
\nonumber
\\
&&
- \bigg(\frac{\bar{\mathcal{C}}_D}{c_D(\mu)}-\frac{1}{2} \bar{\mathcal{C}}_v \bigg) \frac{\rho_D^3}{2m_b^3}
 - \bigg(\frac{\bar{\mathcal{C}}_{LS} }{c_S(\mu)} - \frac{1}{2} \bar{\mathcal{C}}_v \bigg) \frac{\rho_{LS}^3}{2m_b^3}
 \bigg]
 \\
 &=& \Gamma^0 |V_{qb}|^2 
 \bigg(   \mathcal{C}_0 
- \mathcal{C}_{\mu_\pi}\frac{\mu_\pi^2}{2m_b^2}
+ \mathcal{C}_{\mu_G}\frac{\mu_G^2}{2m_b^2}
- \mathcal{C}_{\rho_D} \frac{\rho_D^3}{2m_b^3}
 - \mathcal{C}_{\rho_{LS}} \frac{\rho_{LS}^3}{2m_b^3}
 \bigg)
  \nonumber
  \, , 
\end{eqnarray}
where $r = q^2/m_b^2$ with phase space constrained to $\eta \le r \le (1-\sqrt{\rho})^2$. 
The coefficients $\bar{\mathcal{C}}_i\equiv \mathcal{C}_i - \mathcal{C}_0 \tilde{C}_i$ are defined as the difference 
between the coefficients $\mathcal{C}_i$ of the HQE of the transition operator (in differential form) and the current Eq.~(\ref{hqebvb}) multiplied 
by $\mathcal{C}_0$. Note that the $q^2$ spectrum also satisfies reparametrization invariance, so 
$\mathcal{C}_0 = \mathcal{C}_{\mu_\pi}$ and $\mathcal{C}_{\mu_G} = \mathcal{C}_{\rho_{LS}}$~\cite{Fael:2018vsp}.

The coefficients $\mathcal{C}_i$ ($i=0,\,\mu_\pi,\,\mu_G,\,\rho_D,\,\rho_{LS}$) of the differential rate depend on
three ratios $r$, $\rho$ and $\eta$ and they are related to the corresponding coefficients $C_i$ of the total rate
by
\begin{eqnarray}
 C_i(\rho,\eta) &=& \int_{\eta}^{(1-\sqrt{\rho})^2}dr\, \mathcal{C}_i(r,\rho,\eta)\,.
 \label{Citot}
\end{eqnarray}
However, as it happens with the lepton energy spectrum, the $q^2$ spectrum cannot be interpreted point-by-point. 
In order to compare to experiment, one computes moments of the spectrum for which a clean HQE exists. 
In analogy to the total rate, the HQE for $q^2$ moments is written as

\begin{eqnarray}
 \label{hqemoments}
M_n(B\rightarrow X_q \tau \bar{\nu}_\tau ) &=& \int_{\eta}^{(1-\sqrt{\rho})^2}dr\, r^n \frac{d\Gamma(B\rightarrow X_q \tau \bar{\nu}_\tau )}{dr}
\\
&=&
\Gamma^0 |V_{qb}|^2 
 \bigg[   M_{n,0} 
- M_{n,\mu_\pi}\frac{\mu_\pi^2}{2m_b^2}
+ M_{n,\mu_G}\frac{\mu_G^2}{2m_b^2}
- M_{n,\rho_D} \frac{\rho_D^3}{2m_b^3}
 - M_{n,\rho_{LS}} \frac{\rho_{LS}^3}{2m_b^3}
 \bigg]
  \nonumber
  \, , 
\end{eqnarray}
whose coefficients are related to the coefficients of the differential rate by

\begin{eqnarray}
 M_{n,i}(\rho,\eta) &=& \int_{\eta}^{(1-\sqrt{\rho})^2}dr\, r^n \mathcal{C}_i(r,\rho,\eta)\,.
 \label{Mitot}
\end{eqnarray}
Note that the coefficients $M_{0,i}$ of the zeroth moment correspond the coefficients $C_i$ of the total rate.

Finally, note that the minimum dilepton invariant mass in inclusive semileptonic decays is $q^2_{\rm min} = m_\ell^2$. In the case of decays to 
electron or muon the masses can be safely disregarded and $q^2_{\rm min} = 0$. Such a low $q^2$ is difficult to detect and 
experimentalists use cuts while integrating up to the available $q^2$ \cite{Belle:2021idw}. However, for decays to 
tau $q^2_{\rm min} = m_\tau^2 \sim 3.17$ GeV$^2$ and it might not be necessary to introduce cuts. For example, the lowest $q^2$
cut introduced in the measurement of $B\rightarrow X_c e\bar{\nu}_e$ moments by \cite{Belle:2021idw} was $q^2_{\rm min} = 3.0$ GeV$^2$, which suggests that 
for semitauonic decays the whole $q^2$ range can be measured. 
Despite of this, it might be useful to study moments as a function of low $q^2$ cuts if one pursues an analysis similar to the one 
in~\cite{Bernlochner:2022ucr}. Cut moments can be easily obtained from Eqs.(\ref{hqemoments}) and (\ref{Mitot}) by integrating in the desired range.

\section{Outline of the calculation}
\label{sec:HQE}

This section sketches the outline of the calculation. First one computes the matching coefficients for the $q^2$ spectrum 
by using the spectral representation given in Eq.~(\ref{spectrep}), and latter one integrates over $q^2$ to obtain the coefficients of the 
total rate. Since Eqs. (\ref{eq:HQE-1}), (\ref{hqebvb}) and (\ref{LHQET}) hold a the operator level, the matching calculation can be performed by computing 
on shell matrix elements with partonic states i.e. with quarks and gluons. For power corrections it is possible 
to compute the coefficients by using a small momentum expansion near the heavy quark mass shell~\cite{Mannel:2015jka,Mannel:2019qel}. 

The LO-QCD and NLO-QCD contributions to the differential rate are given by one-loop and two-loop heavy quark to heavy quark ($b\rightarrow b$) or heavy 
quark to gluon heavy quark ($b\rightarrow g b$) scattering Feynman diagrams which are computed in dimensional regularization. The Feynman gauge is chosen.
The scattering is computed in the external gluonic field by using the background field method.  
The amplitudes are reduced to a combination of master integrals by 
using LiteRed~\cite{Lee:2012cn,Lee:2013mka}. The corresponding master integrals were computed analytically in \cite{Mannel:2021ubk}. 
Dirac algebra is handled by using Tracer~\cite{Jamin:1991dp}.

As for renormalization, the $\overline{\mbox{MS}}$ scheme is used for the strong coupling $\alpha_s(\mu)$ and the HQET Lagrangian. 
The bottom and charm quarks are renormalized on-shell. That is 

\begin{eqnarray}
 b_B = (Z_2^{\mbox{\scriptsize OS}})^{1/2} b\,,\quad\quad  m_{c,B} = Z_{m_c}^{\mbox{\scriptsize OS}} m_c
 \,,\quad\quad 
 Z_{m_q}^{\mbox{\scriptsize OS}} &=& 1 - C_F \frac{\alpha_s(\mu)}{4\pi}\bigg( \frac{3}{\epsilon} + 6 \ln\left(\frac{\mu}{m_q}\right) + 4 \bigg)\,,
\end{eqnarray}
where quantities with the subscript $B$ stand for bare, quantities which do not have subscript stand for renormalized, and 
$Z_2^{\mbox{\scriptsize OS}}= Z_{m_b}^{\mbox{\scriptsize OS}}$ to this order. 
Therefore, results are conveniently presented in the the on-shell scheme for both quark masses $m_c$ and $m_b$.

Note that in order to achieve more precise theoretical predictions one typically uses a short distance mass for the bottom quark, like 
the $1S$ mass~\cite{Hoang:1999us,Hoang:1999zc,Hoang:1998hm,Hoang:1998ng,Bauer:2004ve} or the kinetic mass~\cite{Bigi:1994ga,Bigi:1996si}. For the charm 
quark mass one often uses the $\overline{\mbox{MS}}$ mass. 
A change in the mass scheme can be implemented by using the known one-loop relation between the different mass schemes.

For the presentation of results, the coefficients of the $q^2$ spectrum and the total rate are split as it follows ($i = 0,v, \mu_G, \rho_D$)

\begin{eqnarray}   
 \mathcal{C}_i(r,\rho,\eta) &=& \mathcal{C}_i^{\mbox{\scriptsize LO}} 
 + \frac{\alpha_s}{\pi}\bigg(C_F \mathcal{C}_i^{\mbox{\scriptsize NLO, F}}
 + C_A \mathcal{C}_i^{\mbox{\scriptsize NLO, A}}\bigg)\,,
 \\
  C_i(\rho,\eta) &=& C_i^{\mbox{\scriptsize LO}} 
 + \frac{\alpha_s}{\pi}\bigg(C_F C_i^{\mbox{\scriptsize NLO, F}}
 + C_A C_i^{\mbox{\scriptsize NLO, A}}\bigg)\,.
\end{eqnarray}
Analytical results for coefficients of the right-hand side are provided in the Mathematica notebook ``cobqtv.nb'', which is 
supplied as an ancillary file. Following \cite{Mannel:2021ubk}, the coefficients of the $q^2$ spectrum are given in terms of the 
convenient variables
\begin{eqnarray}
	\label{xminus}
	x_{-} &=& \frac{1}{2}\bigg(1-r+\rho - \sqrt{(1-(\sqrt{r}-\sqrt{\rho})^2)(1-(\sqrt{r}+\sqrt{\rho})^2)}\bigg)\,,
	\\
	x_{+} &=& \frac{1}{2}\bigg(1-r+\rho + \sqrt{(1-(\sqrt{r}-\sqrt{\rho})^2)(1-(\sqrt{r}+\sqrt{\rho})^2)}\bigg)\,,
	\label{xplus}
\end{eqnarray}
together with $\eta$. 
The coefficients of the differential rate have a simple dependence on the tau lepton mass. They 
are polynomials of degree three in $\eta$, which follows from Eq.(\ref{spectrep}). In other words, it 
is a consequence of the factorization of the leptonic and hadronic tensors.
Contrarily, the dependence on the the charm quark mass and the invariant mass of the leptons is more complicated and it is efficiently expressed in terms 
of $x_{\pm}$. Analytical results require the use of dilogarithms $\Li_2(x)$.

The coefficients of the total rate are obtained by analytical integration of Eq.~(\ref{Citot}) which, as pointed 
out in~\cite{Mannel:2021ubk}, it can be achieved after changing variables to $x_{-}$ by using $x_{+} = \rho/x_{-}$. 
After the change of variables, the integral in Eq.~(\ref{Citot}) becomes 

\begin{equation}
 C_i(\rho,\eta) = \int_{x_{-}(r=\eta)}^{\sqrt{\rho}}dx_{-} \bigg(\frac{\rho}{x_{-}^2} -1 \bigg) \mathcal{C}_i\left(x_{+} =\frac{\rho}{x_{-}},x_{-},\eta\right)\,.
\end{equation}
The coefficients of the total rate have a complicated dependence on $\rho$, $\eta$. It is convenient to express them in terms of the 
variables $z_{\pm}$ defined by

\begin{eqnarray}
	\label{zminus}
	z_{-} = x_{-}(r=\eta) &=& \frac{1}{2}\bigg(1-\eta+\rho - \sqrt{(1-(\sqrt{\eta}-\sqrt{\rho})^2)(1-(\sqrt{\eta}+\sqrt{\rho})^2)}\bigg)\,,
	\\
	z_{+} = x_{+}(r=\eta) &=& \frac{1}{2}\bigg(1-\eta+\rho + \sqrt{(1-(\sqrt{\eta}-\sqrt{\rho})^2)(1-(\sqrt{\eta}+\sqrt{\rho})^2)}\bigg)\,,
	\label{zplus}
\end{eqnarray}
since they simplify the results drastically. Analytical results require the use of dilogarithms $\Li_2(x)$ and trilogarithms $\Li_3(x)$. 

It is instructive to see how results in the $B\rightarrow X_c \tau \bar{\nu}_\tau$ decay channel can be compactly 
written in terms of $x_{\pm}$ and $\eta$ for the differential rate, and in terms of $z_{\pm}$ for the total rate. 
For this reason, the LO expressions will be explicitly written in the text. The NLO expressions are too lengthy to be explicitly 
written and they are provided only in the ancillary file. For the $B\rightarrow X_u \tau \bar{\nu}_\tau$ decay channel expressions are reasonable in 
length and they will be explicitly shown up to NLO.

\subsection{Leading power coefficient}
\label{difratemb0}

To leading power, the heavy hadron decay corresponds to the free heavy quark decay. 
The NLO QCD corrections to the inclusive $b \rightarrow q \tau \bar{\nu}_\tau$ decay rate and distributions 
were computed analytically to this order almost thirty years ago~\cite{Czarnecki:1994bn,Jezabek:1996db}. 
This section reproduces the known results for the decay rate and the $q^2$ spectrum as a check of the calculation.

\begin{figure}[!htb]  
	\centering
	\includegraphics[width=0.95\textwidth]{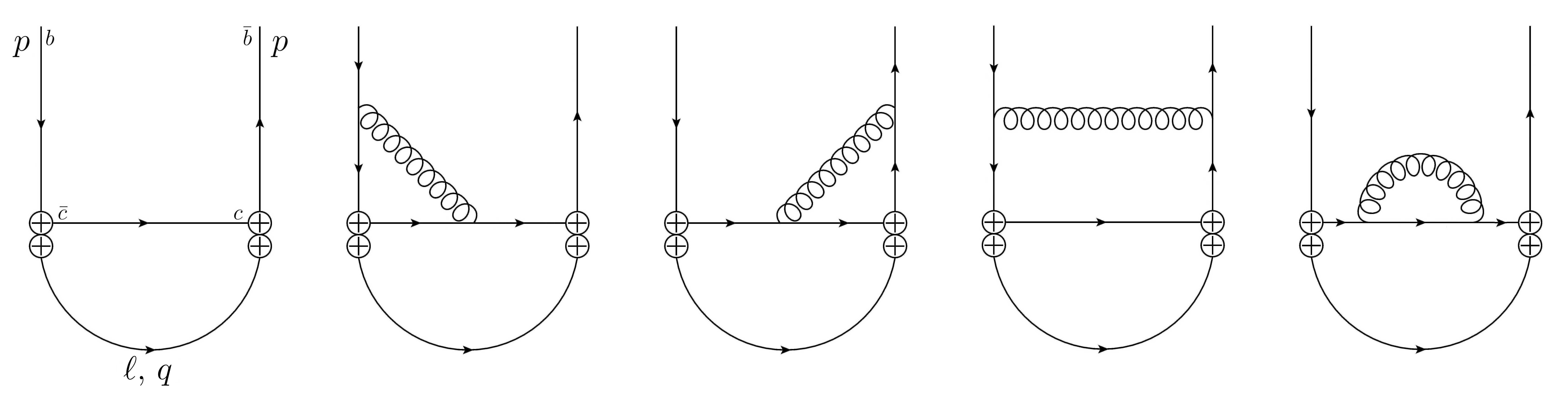}
        \caption{Heavy quark to heavy quark scattering diagrams contributing to the leading power coefficient 
        $\mathcal{C}_0$ in the HQE of the $B \to X_q \tau \bar{\nu}_\tau$ decay spectrum, Eq.~(\ref{hqedifwidth}). 
        Circles with crosses stand for insertions of $\mathcal{L}_{\rm eff}$.
        }
        \label{SampleFDC0NLO}
\end{figure}

For the computation, one takes the heavy quark to heavy quark scattering amplitude with on-shell external momentum $p^2 = m_b^2$ and 
projects it to the $\mathcal{O}_0$ operator. The diagrams that contribute to the coefficient of the differential rate 
at leading power are shown in Fig.~[\ref{SampleFDC0NLO}].
The coefficient of the partonic rate renormalizes through $Z_2^{\rm OS}$ and $Z_{m_c}^{\mbox{\scriptsize OS}}$

\begin{eqnarray}
 \mathcal{C}_0 &= Z_2^{\rm OS} \mathcal{C}_{0,B}(m_{c,B} = Z_{m_c}^{\mbox{\scriptsize OS}} m_c)\,.
\end{eqnarray}
For the $B \rightarrow X_c \tau\bar{\nu}_\tau$ case, the partonic coefficient of the differential rate at LO reads

\begin{eqnarray}
 \mathcal{C}_0^{\rm LO}(r,\rho,\eta) &=& 
 -\frac{2 (x_{-}-x_{+}) (\eta -(1-x_{-})(1-x_{+}) )^2}{(x_{-}-1)^3 (x_{+}-1)^3}
 \bigg[
   \eta  \Big(x_{-} x_{+} (3 x_{-}+3x_{+}-10) + (3-x_{-}) x_{-} 
   \nonumber
   \\
   &&
   - (x_{+}-3) x_{+}\Big) 
   + (x_{-}-1) (x_{+}-1) \Big(x_{-} x_{+} (3 x_{-} + 3x_{+}-8) + (3-2 x_{-}) x_{-} 
   \nonumber
   \\
   &&
   +(3-2 x_{+}) x_{+}\Big)
   \bigg]\,.
   \label{C0LOq2bc}
\end{eqnarray}
At NLO the expression is too lengthy to be explicitly written in the text and it is provided in the ancillary file. 

After integration over $r$ in the whole range one obtains the partonic coefficient of the total rate at NLO. The LO expression 
is obtained after integrating Eq.(\ref{C0LOq2bc}). It reads

\begin{eqnarray}
 C_0^{\rm LO}(\rho,\eta) &=& 
   \left( -1 + 7\rho + 7\rho^2 - \rho^3 + \eta (7 \rho ^2-12 \rho +7) + 7\eta^2 (\rho +1) -\eta^3 \right) (z_{-}-z_{+})
   \nonumber
   \\
   &&   
   + 12 \eta^2 \left(\rho ^2-1\right) \ln \left(\frac{z_{+}-1}{z_{-}-1}\right)
   + 12 \left(\eta ^2-1\right) \rho ^2 \ln\left(\frac{z_{-}}{z_{+}}\right)\,.
\end{eqnarray}
Note that, in the expression above, the overall functions multiplying the $(z_{-}-z_{+})$ and logarithmic structures are written in terms of 
$\rho$ and $\eta$ instead of $z_{\pm}$. To LO this compacts the results significantly. However, 
this is not the case to NLO where writing the expression in terms of $z_{\pm}$ simplifies the result drastically. The coefficients 
given in the ancillary file are written only in terms of $z_{\pm}$. 
Again the NLO the expression is too lengthy to be explicitly written in the text and it is provided in the ancillary file. 

For the leading power coefficient the limit $\rho=0$ can be taken, which allows to find expressions for the 
$B \rightarrow X_u \tau\bar{\nu}_\tau$ case. The coefficient of the differential rate reads

\begin{eqnarray}
\label{C0LOq2bu}
\mathcal{C}_0^{\rm LO}(r,0,\eta) &=& \frac{2 (r-1)^2 (r-\eta )^2}{r^3}\Big(\eta  (r+2)+r (2 r+1)\Big)\,,
\\
 \mathcal{C}_0^{\rm NLO}(r,0,\eta) &=& - \frac{(r-\eta )^2}{6 r^4} \bigg(
    -3\left(6 r^3-15 r^2+4 r+5\right) r^2 
    -3 \eta  \left(3 r^3+6 r^2-25 r+16\right) r
 \nonumber
 \\
 &&
   +8 \pi ^2 \left(\eta  \left(r^3-3 r+2\right) r+(r-1)^2 (2 r+1) r^2\right)  
 \nonumber
 \\
 &&
  -12 \left(\eta  \left(r^2+2 r-2\right) r^2+\left(2r^2+r-1\right) r^3\right) \ln(r)  
  \nonumber
 \\
 &&
 +6  (r-1)^2\left(\eta \left(2 r^2+10 r-3\right)+r^2 (4 r+5) \right) \ln (1-r)
 \nonumber
 \\
 &&
  -12 \left(\eta  \left(r^3-3 r+2\right) r+(r-1)^2 (2 r+1) r^2\right) \ln (1-r) \ln (r)
 \nonumber
 \\
 &&
 -24 \Li_2(1-r) \left(\eta  \left(r^3-3 r+2\right) r+(r-1)^2 (2 r+1) r^2\right)
   \bigg)\,.
   \label{C0NLOq2bu}
\end{eqnarray}
After integration of Eqs. (\ref{C0LOq2bu}) and (\ref{C0NLOq2bu}) over $r$ in the whole range one obtain the partonic coefficient of the total rate. 
It can be also obtained by taking the limit $\rho=0$ in the coefficient of the partonic rate for 
the $B \rightarrow X_c \tau\bar{\nu}_\tau$ case. It reads

\begin{eqnarray}
 C_0^{\rm LO}(0,\eta) &=& 1 - 8\eta + 8\eta^3 - \eta^4  - 12\eta^2 \ln(\eta)\,,
 \\
 C_0^{\rm NLO}(0,\eta) &=& 
  \frac{1}{24} \left(-18 \eta^4+494 \eta ^3+63 \eta ^2-614 \eta +75\right)
 +\frac{1}{6} \pi ^2 \left(2 \eta ^4-16 \eta ^3+36 \eta ^2+24 \eta-3\right)
  \nonumber
 \\
 &&
 + \frac{1}{12} \left(31 \eta ^4-320 \eta ^3+320 \eta -31\right) \ln (1-\eta ) 
 - \frac{1}{12} \left(31 \eta ^3-188 \eta ^2+90 \eta +12\right) \eta  \ln (\eta )
  \nonumber
 \\
 &&
 + \left(\eta ^4-8 \eta ^3+8 \eta -1\right) \ln (1-\eta) \ln (\eta )
 +\left(2 \eta ^4-16 \eta ^3-36 \eta ^2+8 \eta -1\right) \Li_2(\eta )
 \nonumber
 \\
 &&
   +4 \eta ^2 \left(\pi ^2 \ln (\eta )+9
   \Li_3(\eta )-3 \ln (\eta ) \Li_2(\eta )-9 \zeta (3)\right)\,,
\end{eqnarray}
where $\zeta (x)$ is the Riemann zeta function. These expressions are in agreement with the known results~\cite{Czarnecki:1994bn,Bagan:1994zd}. 

By taking the limit $\eta=0$ one recovers the known expressions for the $B \rightarrow X_q e \bar{\nu}_e$ case, where 
leptons are massless~\cite{Nir:1989rm,Mannel:2021ubk,Mannel:2021zzr}.
These two independent limits are a strong check of the calculation.

If desired, the leading power coefficient of moments can be easily obtained by numerical integration of Eq.(\ref{Mitot}).

\subsection{Chromomagnetic operator coefficient}
\label{difratemb2}
The NLO corrections to the $1/m_b^2$ terms in inclusive semileptonic decays of massless leptons are 
known~\cite{Becher:2007tk,Alberti:2012dn,Alberti:2013kxa,Mannel:2014xza,Mannel:2015wsa,Mannel:2015jka,Mannel:2021zzr}, but to the best of my knowledge 
they have never been computed for the case where one of the leptons is massive. However, to LO the $1/m_b^2$ corrections to inclusive semitauonic 
decays have been extensively studied~\cite{Koyrakh:1993pq,Balk:1993sz,Falk:1994gw}. This section addresses the computation of the missing 
$\alpha_s/m_b^2$ corrections for the coefficient of the dilepton invariant mass spectrum and the total width.

\begin{figure}[!htb]  
	\centering
	\includegraphics[width=1.0\textwidth]{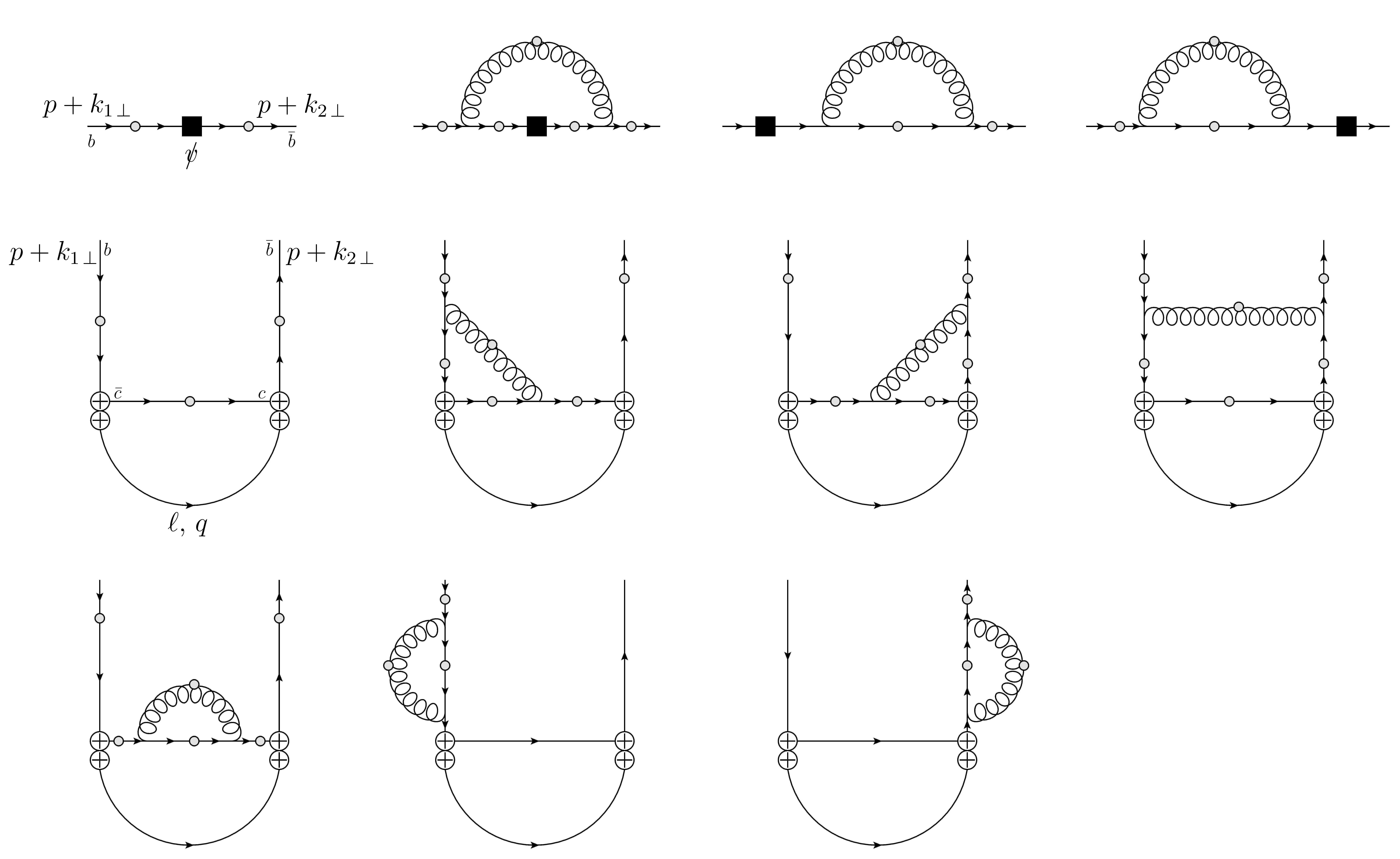}
        \caption{Quark to quark-gluon scattering diagrams contributing to the coefficients 
        $\bar{\mathcal{C}}_i = \mathcal{C}_i - \mathcal{C}_0 \tilde{C}_i$ 
        of power corrections in the HQE of the $B \to X_q \tau \bar{\nu}_\tau$ decay spectrum, Eq.~(\ref{hqedifwidth}). 
        The momentum $p$ is fixed to be on shell, i.e. $p^2=m_b^2$, whereas $k_{i\,\perp}$ are soft momenta. Black squares stand for $\slashed v$ insertions, 
        circles with crosses for insertions of $\mathcal{L}_{\rm eff}$, and gray dots stand for possible gluon insertions 
        with incoming momentum $k_{2\,\perp}-k_{1\,\perp}$. After properly accounting for all one gluon insertions, there are five diagrams at LO-QCD and 
        forty-one diagrams at NLO-QCD.
        }
        \label{SampleFDdifwidth}
\end{figure}

Before addressing the calculation of the $1/m_b^2$ corrections, one needs to compute the auxiliary coefficient $\bar{\mathcal{C}}_v$, which 
appears in the HQE of the transition operator at order $1/m_b$. This coefficient contributes to higher orders in the $1/m_b$ expansion after using the EOM 
and, in particular, shifts 
the coefficients of the $1/m_b^2$ and $1/m_b^3$ terms. For its calculation one takes the amplitude 
of the quark to quark-gluon scattering diagrams given in Fig.~[\ref{SampleFDdifwidth}] with 
vanishing soft momenta $k_{1\,\perp} = k_{2\,\perp}=0$ and
projects it to the $\mathcal{O}_v$ operator. This is achieved by taking a longitudinally polarized gluon 
exchange ($v\cdot\epsilon$) without momentum transfer.
In practice, one directly computes the difference between the HQE of the transition operator and the current
\begin{eqnarray}
 \bar{\mathcal{C}}_v &\equiv& \mathcal{C}_v - \mathcal{C}_0 \tilde{C}_v
 = Z_2^{\mbox{\scriptsize OS}}(\mathcal{C}_{v,B}
 - \mathcal{C}_{0,B} \tilde{C}_{v,B})\,,
\end{eqnarray}
with $\mathcal{C}_{i,B}=\mathcal{C}_{i,B}(m_{c,B} = Z_{m_c}^{\mbox{\scriptsize OS}} m_c)$. The explicit expression for 
the $\bar{\mathcal{C}}_v$ coefficient is not written here because it appears only at intermediate steps. However, the 
analytical expression is provided in the ancillary file.

To order $1/m_b^2$ one only needs to compute the coefficient of $\mu_G^2$, since the coefficient of $\mu_\pi^2$ is related to the 
leading power coefficient due to reparametrization invariance~\cite{Luke:1992cs,Manohar:2010sf}. 
Note that at dimension five the operator 
$\mathcal{O}_{{\scriptsize\mbox{I}}} = \bar h_v (v\cdot\pi)^2 h_v $ also appears,
but it contributes to higher orders in the $1/m_b$ expansion after using the EOM. 
The computation follows the lines of~\cite{Mannel:2021zzr}, where one takes the amplitude of quark to quark-gluon scattering, 
expands to linear order in the small momentum and
projects it to the chromomagnetic operator $\mathcal{O}_G$.

The diagrams that contribute are shown in Fig.~[\ref{SampleFDdifwidth}]. 
For the determination of the chromomagnetic operator coefficient one takes $k_{1\,\perp} = 0$ and $k_{2\,\perp}= k_\perp$. In other words, 
one considers a single small gluon momentum
$k_\perp$. By picking up the structure $[\slashed k_\perp, \slashed \epsilon_\perp]$, where $\epsilon$ is the gluon polarization vector, one 
obtains the desired coefficient. 

Like in the previous case, one directly computes the difference between the HQE of the transition operator and the current

\begin{eqnarray}
 \bar{\mathcal{C}}_G &\equiv& \mathcal{C}_G - \mathcal{C}_0 \tilde{C}_G
 = Z_2^{\mbox{\scriptsize OS}}Z_{\mathcal{O}_G}(\mathcal{C}_{G,B}
 - \mathcal{C}_{0,B}  \tilde{C}_{G,B})\,,
\end{eqnarray}
again with with $\mathcal{C}_{i,B}=\mathcal{C}_{i,B}(m_{c,B} = Z_{m_c}^{\mbox{\scriptsize OS}} m_c)$ and where 

\begin{equation}
 Z_{\mathcal{O}_G} = 1 - C_A \frac{\alpha_s}{4\pi} \frac{1}{\epsilon}\,,
\end{equation}
is the renormalization factor of the chromomagnetic operator. The
coefficient $\bar{\mathcal{C}}_G$ is finite and the cancellation of poles provides a solid check of the computation.
The coefficient of the differential rate $\mathcal{C}_{\mu_G}$ in front of $\mu_G^2$ is finally obtained from

\begin{equation}
 \mathcal{C}_{\mu_G} = \frac{\bar{\mathcal{C}}_G}{c_F(\mu)} - \bar{\mathcal{C}}_v\,.
\end{equation}
For the $B \rightarrow X_c \tau\bar{\nu}_\tau$ case, the chromomagnetic operator coefficient of the differential rate at LO reads

\begin{eqnarray}
 \mathcal{C}_{\mu_G}^{\rm LO}(r,\rho,\eta) &=& - \frac{2 (\eta - (1-x_{+})(1-x_{-}) )^2 }{(x_{-}-1)^3 (x_{+}-1)^3 (x_{-}-x_{+})}
 \bigg[ \eta  \Big(-3 x_{-}^2 x_{+}^2 (x_{-}+x_{+}-6) 
    \nonumber
   \\
   &&
  + x_{-} x_{+} \left(15 x_{-}^3-44 x_{-}^2+33 x_{-}+15 x_{+}^3-44 x_{+}^2+33x_{+}-24\right)
   \nonumber
   \\
   &&
   + x_{-} \left(-5 x_{-}^3+15 x_{-}^2-12 x_{-}+4\right) 
   + x_{+} \left(-5 x_{+}^3+15 x_{+}^2-12x_{+}+4\right) \Big)
   \nonumber
   \\
   &&
   +(x_{-}-1) (x_{+}-1) \Big( - 3 x_{-}^2 x_{+}^2(x_{-}+x_{+}-4)
   \nonumber
   \\
   &&
   + x_{-} x_{+} \left( 15 x_{-}^3 + 15 x_{+}^3 - 28 x_{-}^2 - 28 x_{+}^2 + 9 x_{-} + 9x_{+} \right)
   \nonumber
   \\
   &&
  + x_{-}(- 10x_{-}^3 + 15x_{-}^2 - 4)
  + x_{+}(- 10x_{+}^3 + 15x_{+}^2 - 4)
  \Big) \bigg]\,.
  \label{CmuGLOq2bc}
\end{eqnarray}
At NLO the expression is too lengthy to be explicitly written in the text and it is provided in the ancillary file. 

After integration over $r$ in the whole range one obtains the chromomagnetic operator coefficient of the total rate at NLO. 
The LO expression is obtained after integrating Eq.(\ref{CmuGLOq2bc}). It reads

\begin{eqnarray}
 C_{\mu_G}^{\rm LO}(\rho,\eta) &=&
   \left(  3 -5 \rho + 19\rho^2 - 5\rho^3 
   + \eta\left(35 \rho ^2-28 \rho -5\right) + \eta^2 (35 \rho +19) - 5\eta^3
   \right) (z_{-}-z_{+})
   \nonumber
   \\
   &&
   + 12 \eta^2 \left(5 \rho ^2-1\right) \ln \left(\frac{z_{+}-1}{z_{-}-1}\right) 
   + 12 \left(5 \eta ^2-1\right) \rho ^2 \ln\left(\frac{z_{-}}{z_{+}}\right)\,.
\end{eqnarray}
Again the NLO the expression is too lengthy to be explicitly written in the text and it is provided in the ancillary file. 

For the chromomagnetic operator coefficient the limit $\rho=0$ can be taken, which allows to find expressions for the 
$B \rightarrow X_u \tau\bar{\nu}_\tau$ case. The coefficient of the differential rate reads

\begin{eqnarray}
 \mathcal{C}_{\mu_G}^{\rm LO}(r,0,\eta) &=& \frac{2 (r-\eta )^2}{r^3}\Big(10 r^4 -15 r^3+r + \eta(5 r^3-3 r+2)\Big) \,,
 \label{CmuGLOq2bu}
 \\
 \mathcal{C}_{\mu_G}^{\rm NLO,\, F}(r,0,\eta) &=& - \frac{(r-\eta )^2}{18 r^4} \bigg[
    3\left(-138 r^3+313 r^2-212 r+5\right) r^2 +3 \eta  \left(-69 r^3-82 r^2+95 r+16\right) r
   \nonumber
   \\
   &&
   +8 \pi ^2 \left(\eta  \left(15 r^3-4 r^2-5 r-2\right) r+\left(30r^3-53 r^2+20 r-1\right) r^2\right)
   \nonumber
   \\
   &&
   -12 \left(\eta  \left(5 r^2+6 r+2\right) r^2+\left(10 r^2-3 r+1\right) r^3\right) \ln(r)
   \nonumber
   \\
   &&
   +6 \Big(\eta  \left(10 r^4+102 r^3-115 r^2+60 r-21\right)
   \nonumber
   \\
   &&
   +r \left(20 r^4+39 r^3-110r^2+27 r-12\right)\Big) \ln (1-r) 
   \nonumber
   \\
   &&
   +12 \left(\eta  \left(-15 r^3+4 r^2+5 r+2\right) r+\left(-30 r^3+53 r^2-20 r+1\right) r^2\right) \ln(1-r) \ln(r)  
   \nonumber
   \\
   &&
   -24 \Li_2(1-r) \left(r^2 \left(30 r^3-53 r^2+20 r-1\right)-\eta  r \left(-15 r^3+4 r^2+5r+2\right) \right)
   \bigg]\,,
   \label{CmuGNLOCFq2bu}
   \\
    \mathcal{C}_{\mu_G}^{\rm NLO,\, A}(r,0,\eta) &=& - \frac{(r-\eta )^2}{18 r^3} 
 \bigg[ 
   3 r \left(90 r^3-99 r^2-90 r+23\right) + 3\eta  \left(45 r^3+18 r^2-9 r+22\right) 
   \nonumber
   \\
   &&
   -8 \pi ^2 \left(\eta  \left(4 r^2-5 r+2\right)+r \left(8 r^2-10 r+1\right)\right)
   \nonumber
   \\
   &&
   +6 \left(\eta  \left(3 r^2+8 r-4\right) r+\left(6 r^2+7r-2\right) r^2\right) \ln(r)
   \nonumber
   \\
   &&
   + 6 \left(-6r^4 + 17r^3 - 44r^2 + 15r + 6 -\eta\left(3 r^3-4r^2+r-12\right) \right) \ln (1-r)
   \nonumber
   \\
   &&   
   +12 \left(\eta  \left(4 r^2-5 r+2\right)+r \left(8 r^2-10 r+1\right)\right) \ln(1-r) \ln(r)
   \nonumber
   \\
   &&
   + 24 \Li_2(1-r) \left(\eta  \left(4 r^2-5 r+2\right)+r \left(8 r^2-10 r+1\right)\right)
   \bigg]\,.
   \label{CmuGNLOCAq2bu}
\end{eqnarray}
After integration of Eqs. (\ref{CmuGLOq2bu}), (\ref{CmuGNLOCFq2bu}) and (\ref{CmuGNLOCAq2bu}) over $r$ in the whole range one obtains 
the chromomagnetic operator coefficient of the total rate. 
It can be also obtained by taking the limit $\rho=0$ in the coefficient of the chromomagnetic operator of the rate for 
the $B \rightarrow X_c \tau\bar{\nu}_\tau$ case. It reads

\begin{eqnarray}
C_{\mu_G}^{\rm LO}(0,\eta) &=& -3 + 8\eta - 24\eta^2 + 24\eta^3 -5\eta^4 - 12\eta^2 \ln(\eta)\,,
\\
C_{\mu_G}^{\rm NLO,\, F}(0,\eta) &=& 
    \frac{1}{216} \left(-1242 \eta^4+22206\eta^3-19805 \eta ^2-886 \eta -273\right)
      \nonumber
   \\
   &&
   +\frac{1}{54} \pi^2 \left(90 \eta ^4-212 \eta ^3+864 \eta ^2+144 \eta-15\right)
   \nonumber
   \\
   &&
   +\frac{1}{108} \left(1035 \eta ^4-11224 \eta ^3+8856 \eta ^2+1368 \eta -35\right) \ln (1-\eta )
   \nonumber
   \\
   &&
   + \frac{1}{108} \left(-1035 \eta ^3+4348 \eta ^2-3522 \eta +84\right) \eta  \ln (\eta )
      \nonumber
   \\
   &&
   +\frac{1}{9} \left(45
   \eta ^4-196 \eta ^3+108 \eta ^2+36 \eta +7\right) \ln (1-\eta ) \ln (\eta )
      \nonumber
   \\
   &&
   +\frac{1}{9} \left(90 \eta ^4-572 \eta ^3-432 \eta^2+43\right) \Li_2(\eta )
      \nonumber
   \\
   &&
   -\frac{4}{9} (2 \eta -15) \eta ^2 \left(\pi ^2 \ln (\eta )+9 \Li_3(\eta )-3 \ln (\eta ) \Li_2(\eta )-9 \zeta (3)\right)\,,
\\
C_{\mu_G}^{\rm NLO,\, A}(0,\eta) &=& 
   \frac{1}{108} \left(405\eta ^4-1986 \eta ^3+2537 \eta ^2-1142 \eta +186\right)
   \nonumber
   \\
   &&
   + \frac{1}{27} \pi ^2 \left(62 \eta ^3-126 \eta ^2+36 \eta -3\right) 
    \nonumber
   \\
   &&
   + \frac{1}{54} \left(-27 \eta ^4+52 \eta ^3+540 \eta ^2-756 \eta +191\right) \ln (1-\eta )
   \nonumber
   \\
   &&
   + \frac{1}{54} \left(27 \eta ^3-16 \eta ^2-132 \eta -48\right)\eta \ln (\eta )
   \nonumber
   \\
   &&
   +\frac{4}{9} \left(11 \eta ^3-18 \eta ^2+9 \eta -2\right) \ln (1-\eta ) \ln (\eta )  
   \nonumber
   \\
   &&
   +\frac{2}{9} \left(26 \eta ^3-18 \eta ^2+36 \eta -13\right) \Li_2(\eta )
   \nonumber
   \\
   &&
   -\frac{8}{9} \eta ^3
   \left(\pi ^2 \ln (\eta )+9 \Li_3(\eta )-3 \ln (\eta ) \Li_2(\eta )-9 \zeta (3)\right)\,.
\end{eqnarray}
To the best of my knowledge, the $\alpha_s/m_b^2$ results for the $q^2$ spectrum and the total rate for
both decay channels $B \rightarrow X_c \tau \bar{\nu}_\tau$ and 
$B \rightarrow X_u \tau \bar{\nu}_\tau$ are new. To LO the results agree with~\cite{Ligeti:2014kia,Mannel:2017jfk}. 

For $\eta=0$ one recovers the known results for the chromomagnetic operator coefficient of the leptonic invariant mass spectrum~\cite{Mannel:2021zzr}
and the total rate~\cite{Mannel:2014xza,Mannel:2015wsa,Mannel:2015jka} in $B \rightarrow X_c e \bar{\nu}_e$, where leptons are massless. 
Note that a typo in the latter references was identified and corrected 
in~\cite{Mannel:2021zzr}.

If desired, the chromomagnetic operator coefficient of moments can be easily obtained by numerical integration of Eq.(\ref{Mitot}).

\subsection{Darwin operator coefficient}
\label{difratemb3}

The $1/m_b^3$ corrections in inclusive semitauonic decays have been considered to LO in~\cite{Mannel:2017jfk,Colangelo:2020vhu,Rahimi:2022vlv}, where 
expressions for the lepton energy spectrum and total rate have been obtained. To the best of my knowledge, they have never been considered for the $q^2$ spectrum. 
The NLO corrections to the $1/m_b^3$ terms in inclusive semileptonic decays of massless leptons have been 
computed quite recently~\cite{Mannel:2021zzr}. However, they are unknown for inclusive semitauonic decays.
This section addresses the computation of the $\alpha_s/m_b^3$ corrections for the coefficient of the dilepton invariant mass spectrum 
and the total rate.

In this case, in addition to the Darwin and spin-orbit operators, there are five more operators which nevertheless 
contribute to higher orders in the $1/m_b$ expansion after using the EOM. The contributions to these 
operators must be properly disentangled from the contributions to the Darwin and spin-orbit terms by choosing appropriate projectors. 
Note that reparametrization invariance relates the coefficient of 
$\rho_{\rm LS}^3$ and $\mu_G^2$, so one only needs to compute the coefficient of $\rho_D^3$.

The computation follows the lines of~\cite{Mannel:2021zzr} where one takes 
the amplitude of quark to quark-gluon scattering, expands to quadratic order in the soft momenta and
projects it to Darwin operator $\mathcal{O}_D$. The diagrams that contribute are shown in Fig.~[\ref{SampleFDdifwidth}]. 
For the determination of the Darwin operator coefficient one takes the external gluon 
to have longitudinal polarization ($v\cdot\epsilon$), use two soft quark momenta $k_{1\,\perp}$ and $k_{2\,\perp}$, 
and pick up the structure $k_{1\,\perp}\cdot k_{2\,\perp}$. 

Like before, one directly computes the difference between the HQE of the transition operator and the current
\begin{eqnarray}
\bar{\mathcal{C}}_D \equiv \mathcal{C}_D - \mathcal{C}_0  \tilde{C}_D 
&=& Z_2^{\mbox{\scriptsize OS}}Z_{\mathcal{O}_D}(\mathcal{C}_{D,B} - \mathcal{C}_{0,B}  \tilde{C}_{D,B})
+ \delta \bar{\mathcal{C}}_{D}^{mix}\,,
\end{eqnarray}
with $\mathcal{C}_{i,B}=\mathcal{C}_{i,B}(m_{c,B} = Z_{m_c}^{\mbox{\scriptsize OS}} m_c)$ and where  
\begin{eqnarray}
 Z_{\mathcal{O}_D} &=& - \frac{1}{6}C_A \frac{\alpha_s}{\pi}\frac{1}{\epsilon}\,,
 \\
\delta \bar{\mathcal{C}}_{D}^{mix} &=& 
  \bigg[
   C_F\bigg( 
     \frac{4}{3}\bar{\mathcal{C}}_\pi
   - \frac{2}{3}\bar{\mathcal{C}}_v
   \bigg) 
 + C_A\bigg(
  \frac{5}{12}\bar{\mathcal{C}}_G  
 + \frac{1}{12}\bar{\mathcal{C}}_\pi 
 - \frac{1}{4}\bar{\mathcal{C}}_v 
 \bigg)
 \bigg]
 \frac{\alpha_s}{\pi}\frac{1}{\epsilon}\,,
 \label{CDmix}
\end{eqnarray}
are the renormalization factor of the Darwin operator and
the contribution to the Darwin coefficient coming 
from the HQET operator mixing under renormalization~\cite{Falk:1990pz,Bauer:1997gs,Finkemeier:1996uu,Balzereit:1996yy,Blok:1996iz,Lee:1991hp,Moreno:2017sgd,Lobregat:2018tmn,Moreno:2018lbo}, 
respectively. The quantity $\bar{\mathcal{C}}_D$ is finite and the cancellation of 
poles provides a solid check of the calculation~\cite{Mannel:2021zzr}.

The coefficient of the differential rate $\mathcal{C}_{\rho_D}$ in front of $\rho_D^3$ is finally obtained from
\begin{equation}
 \mathcal{C}_{\rho_D} = \frac{\bar{\mathcal{C}}_D}{c_D(\mu)} - \frac{1}{2} \bar{\mathcal{C}}_v\,.
\end{equation}
For the $B \rightarrow X_c \tau\bar{\nu}_\tau$ case, the Darwin operator coefficient of the differential rate at LO reads

\begin{eqnarray}
 \mathcal{C}_{\rho_D}^{\rm LO}(r,\rho,\eta) &=& \frac{2 (\eta - (1-x_{-})(1-x_{+}) )^2}{3 (x_{-}-1)^3 (x_{+}-1)^3 (x_{-}-x_{+})^3}
 \bigg[ 
   \eta \Big(-2 x_{-}^3 x_{+}^3 (33 x_{-}+33x_{+}-214)
    \nonumber
 \\
 &&
   +x_{-}^2 x_{+}^2 \left(3 x_{-}^3+101 x_{-}^2-570 x_{-}+3 x_{+}^3+101 x_{+}^2-570x_{+}+672\right)
    \nonumber
 \\
 &&
   +x_{-}^2 \left(-5 x_{-}^4+39 x_{-}^3-96 x_{-}^2+88 x_{-}-48\right)
   +x_{-} x_{+}(15 x_{-}^5-70 x_{-}^4+51 x_{-}^3
       \nonumber
 \\
 &&
   +240 x_{-}^2-328 x_{-}+15 x_{+}^5-70 x_{+}^4+51x_{+}^3+240 x_{+}^2-328 x_{+}+192)
       \nonumber
 \\
 &&
   +\left(-5 x_{+}^4+39 x_{+}^3-96 x_{+}^2+88 x_{+}-48\right)
   x_{+}^2\Big)
 \nonumber
 \\
 &&
   +(x_{-}-1) (x_{+}-1) \Big(-22 x_{-}^3 x_{+}^3 (3 x_{-}+3 x_{+}-16)
       \nonumber
 \\
 &&
   +x_{-}^2 x_{+}^2 \left(3 x_{-}^3+154 x_{-}^2-570 x_{-}+3 x_{+}^3+154 x_{+}^2-570
   x_{+}+720\right)
       \nonumber
 \\
 &&
   +x_{-}^2 \left(-10 x_{-}^4+63 x_{-}^3-120 x_{-}^2+104 x_{-}-48\right)
   +x_{-} x_{+} (15 x_{-}^5-80 x_{-}^4+27 x_{-}^3
       \nonumber
 \\
 &&
   +240 x_{-}^2-344 x_{-}+15 x_{+}^5-80 x_{+}^4+27
   x_{+}^3+240 x_{+}^2-344 x_{+}+192)
      \nonumber
 \\
 &&
  +\left(-10 x_{+}^4+63 x_{+}^3-120 x_{+}^2+104x_{+}-48\right) x_{+}^2\Big)
   \bigg]\,.
   \label{CrhoDLOq2bc}
\end{eqnarray}
At NLO the expression is too lengthy to be explicitly written in the text and it is provided in the ancillary file. 

After integration over $r$ in the whole range one obtains the Darwin operator coefficient of the total rate at NLO. 
The LO expression is obtained after integrating Eq.(\ref{CrhoDLOq2bc}). It reads

\begin{eqnarray}
 C_{\rho_D}^{\rm LO}(\rho,\eta) &=& 
 \frac{1}{3} \left( 77 - 11\rho + 13\rho^2 + 5\rho^3 + \eta \left(-35 \rho ^2-12 \rho +13\right) - \eta^2(35 \rho +59) + 5\eta^3 \right) (z_{-}-z_{+})
 \nonumber
 \\
 &&
 - 4 \eta ^2 \left(4 \eta +5 \rho ^2-1\right) \ln \left(\frac{z_{+}-1}{z_{-}-1}\right) 
 - 4 \left(4 \eta ^3+\eta ^2 \left(5\rho ^2-4\right)-4 \eta +3 \rho ^2+4\right) \ln \left(\frac{z_{-}}{z_{+}}\right)\,.
 \nonumber
 \\
 &&
 \label{CrhoDLObc}
\end{eqnarray}
Again the NLO the expression is too lengthy to be explicitly written in the text and it is provided in the ancillary file. 

The LO expression in Eq.~(\ref{CrhoDLObc}) confirms the result given in \cite{Rahimi:2022vlv} 
and disagrees with~\cite{Mannel:2017jfk,Colangelo:2020vhu}. In the case $\eta=\rho$, Eq.~(\ref{CrhoDLObc}) can be compared to the 
$C_1^2$ structure in the nonleptonic $b\rightarrow c\bar c s$ decay~\cite{Mannel:2020fts,Lenz:2020oce} for which we 
find agreement. To the best of my knowledge, the LO result for the Darwin coefficient of the $q^2$ spectrum has never presented explicitly.
The NLO results are new.

Note that the matching is performed by integrating out the charm quark simultaneously
with the hard modes of the bottom quark. That means $m_c^2/m_b^2$ is treated as a number   
of order one in the limit $m_b \to \infty$, which implies that $m_c \to \infty$ as well. 
Therefore, the results can not be extrapolated to the $\rho=0$ ($B\rightarrow X_u \tau \bar{\nu}_\tau$) case in general. In particular, the limit 
$\rho\rightarrow 0$ happens to be non-singular for power corrections up to $1/m_b^2$ and it can be taken, but this is no longer true for the Darwin term. 
This feature shows up as a logarithmic singularity in the Darwin coefficient when taking the limit $\rho\rightarrow 0$. 
In the calculation for strict $\rho=0$ these singularities show up as poles which point out the mixing under renormalization of the Darwin operator 
with dimension six four quark operators (see e.~g. \cite{Mannel:2020fts,Lenz:2020oce,MorenoTorres:2020xir,Piscopo:2021ogu}). This is an additional complication 
beyond the scope of this work. Therefore, $\alpha_s/m_b^3$ corrections are presented only for the $B\rightarrow X_c \tau \bar{\nu}_\tau$ case.

For $\eta=0$ one recovers the known results for the Darwin coefficient of the leptonic invariant mass spectrum~\cite{Mannel:2021zzr}
and the total rate~\cite{Gremm:1996df,Mannel:2021zzr} in the case of massless leptons $B \rightarrow X_c e \bar{\nu}_e$. 
For the total rate, the coefficient previously computed in~\cite{Mannel:2019qel} was corrected in \cite{Mannel:2021zzr}.

As discussed in~\cite{Mannel:2021zzr}, performing the integration over $r$ of the Darwin coefficient to obtain the coefficients of the 
total rate and moments is rather subtle, since the integral is infrared (IR) singular at the upper integration limit $r_{\rm max} =(1-\sqrt{\rho})^2$, 
whereas the regularization parameter $\epsilon$ has been omitted after renormalization of the differential rate. 
This feature points out that expansion in $\epsilon$ and integration over $r$ are operations which do not commute in general. 
Therefore, computing the coefficient of the total width and moments requires restoring the $\epsilon$ dependence in the IR singular terms. 
For the Darwin term there is a single IR divergent integral with singularity $(r_{\rm max} - r)^{-3/2}$, which should be understood as the following dimensionally regulated integral

\begin{eqnarray}
 \int_\eta^{r_{\rm max}} \frac{dr}{(r_{\rm max} - r)^{3/2}} \quad &\longrightarrow& \quad
   \int_\eta^{r_{\rm max}} \frac{dr}{(r_{\rm max} - r)^{3/2 + \epsilon}}
  =-\frac{2}{\sqrt{r_{\rm max} - \eta }} + \mathcal{O}(\epsilon)\,.
\end{eqnarray}
For the integration one takes $\epsilon$ to be an arbitrary complex number and computes the integral in a domain of convergence. Later on one performs an analytic continuation in $\epsilon$ through the complex plane to $\epsilon=0$. Note that the IR divergent integral is finite in dimensional regularization due to the power $3/2$. As a consequence, there is no
generation of new poles after integration. It is instructive to observe that the IR divergent integral can be interpreted as the following distribution

\begin{eqnarray}
 \int_\eta^{r_{\rm max}} dr f(r,\eta) \frac{1}{(r_{\rm max} - r)^{3/2}} \quad &\longrightarrow& \quad
  \int_\eta^{r_{\rm max}}dr f(r,\eta) \bigg[ \frac{1}{(r_{\rm max} - r)_{+}^{3/2}} -\frac{2}{\sqrt{r_{\rm max} - \eta }}\delta(r - r_{\rm max}) \bigg],
\end{eqnarray}
where $f(r,\eta)$ is an arbitrary test function regular at $r=r_{\rm max}$ and the plus distribution
$(r_{\rm max} - r)_{+}^{3/2}$ is defined as follows

\begin{equation}
 \frac{1}{(r_{\rm max} - r)_{+}^{3/2}} f(r,\eta) \equiv \frac{1}{(r_{\rm max} - r)^{3/2}}[ f(r,\eta) - f(r=r_{\rm max},\eta) ]\,.
\end{equation}
Note the close connection between IR singular integrals at the endpoint, the corresponding 
dimensionally regulated integrals, and delta functions sitting at the endpoint. Note that, in contrast to what it happens in 
the charged lepton energy spectrum, the coefficient function in front of the delta distribution is not singular.

In this case, the computation of the Darwin coefficient of moments 
by numerical integration of Eq.~(\ref{Mitot}) is more involved, since it requires computing integrals in dimensional regularization. 
However, the only integral which needs to be regulated is the total rate, for which analytical 
results are provided. In other words, all moments of the form Eq.~(\ref{Mitot}) can be related to the total rate and IR safe integrals 
that can be evaluated numerically if desired

\begin{eqnarray}
 M_{n,\rho_D} = \int_{\eta}^{r_{\rm max}} dr\, r^n \mathcal{C}_{\rho_D} 
     = r_{\rm max}^n C_{\rho_D} + \sum_{k=1}^{n} r_{\rm max}^{k-1} \int_{\eta}^{r_{\rm max}} dr\, r^{n-k} (r-r_{\rm max})\mathcal{C}_{\rho_D}\,.
       \label{momrhoD}
\end{eqnarray}

\section{Numerical analysis}
\label{sec:disc}

This section presents a numerical analysis in order to illustrate the size of the corrections which have been computed. 
The numerical values summarized in table~\ref{tab:par} are taken for illustration.
\begin{table}
 \begin{center}
\begin{tabular}{|c|c|c|c|}
\hline
 Parameter & Numerical value & Parameter & Numerical value\\ 
   \hline
 $\mu=m_b$ & $4.7$ GeV & $\rho_{LS}^3$& $-0.15$ GeV$^3$\\ 
 $\rho=m_c^2/m_b^2$& $0.077$ &  $r_{\rm min}=\eta$ & $0.140$\\ 
 $\eta=m_\tau^2/m_b^2$& $0.14$ & $r_{\rm max}=(1-\sqrt{\rho})^2$ &  $0.522$\\ 
 $\alpha_s(m_b)$ & $0.215$ & $q^2_{\rm min}= m_b^2 r_{\rm min}$ & $3.09$ GeV$^2$\\
 $\mu_\pi^2$ & $0.4$ GeV$^2$ & $q^2_{\rm max}= m_b^2 r_{\rm max}$ & $11.53$ GeV$^2$\\
 $\mu_G^2$ &  $0.35$ GeV$^2$ & $|V_{cb}|$ & $41\cdot 10^{-3}$\\ 
 $\rho_{D}^3$ & $0.2$ GeV$^3$ & $|V_{ub}|$ & $3.82\cdot 10^{-3}$\\ 
 \hline
\end{tabular}
\caption{Numerical values of parameters used in plots and tables. For the matrix elements the values are taken from~\cite{Benson:2003kp}.}
\label{tab:par}
\end{center}
\end{table}

\begin{figure}[ht]
\centering
\subfigure[Leading power coefficient.]{\includegraphics[scale=0.615]{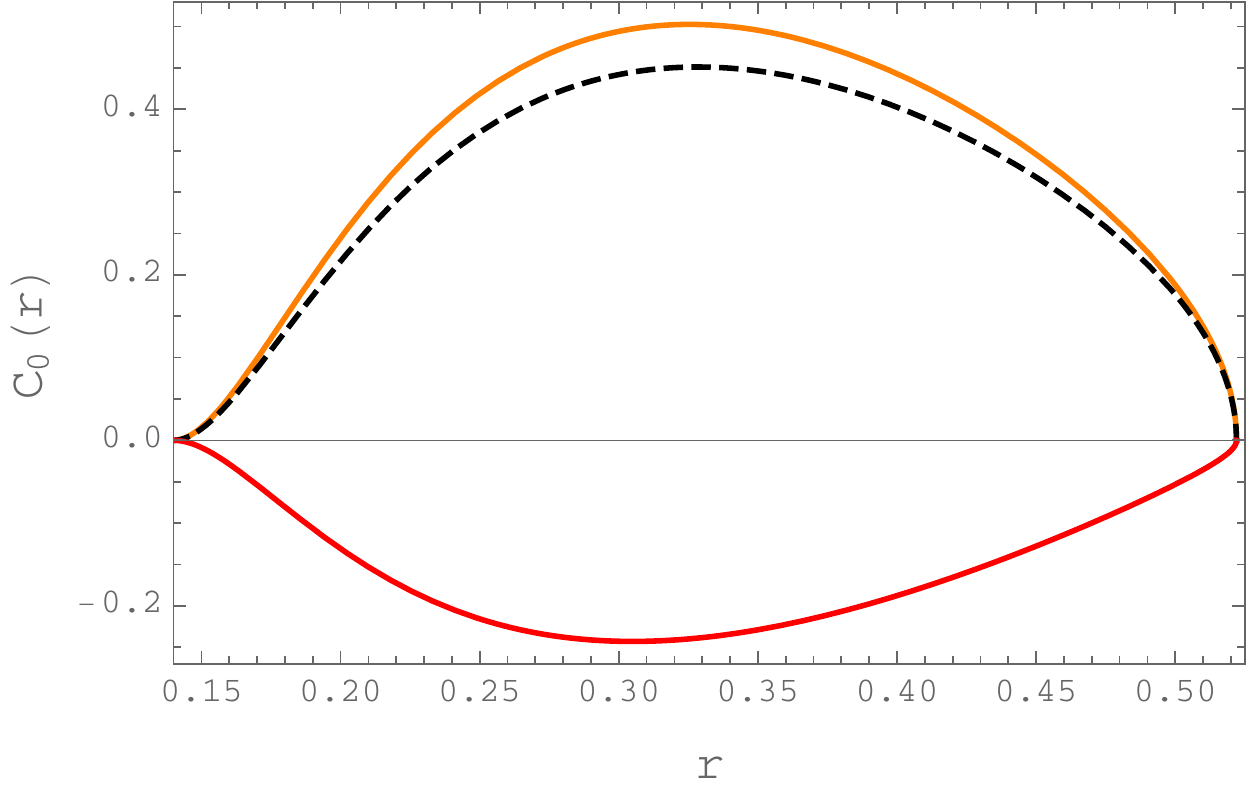}
\label{fig:subfigure1}}
\quad
\subfigure[Chromomagnetic operator coefficient.]{%
\includegraphics[scale=0.6]{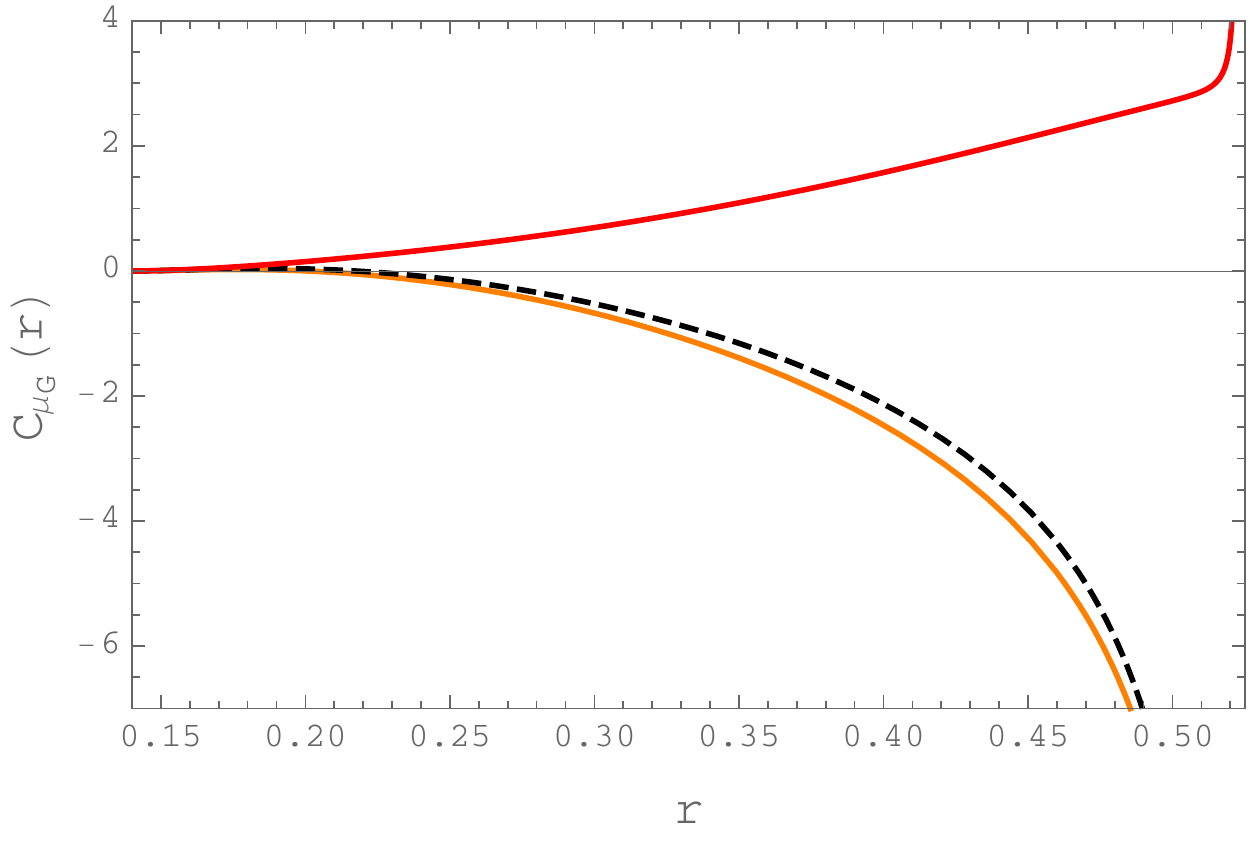}
\label{fig:subfigure3}}
\subfigure[Darwin operator coefficient.]{%
\includegraphics[scale=0.6]{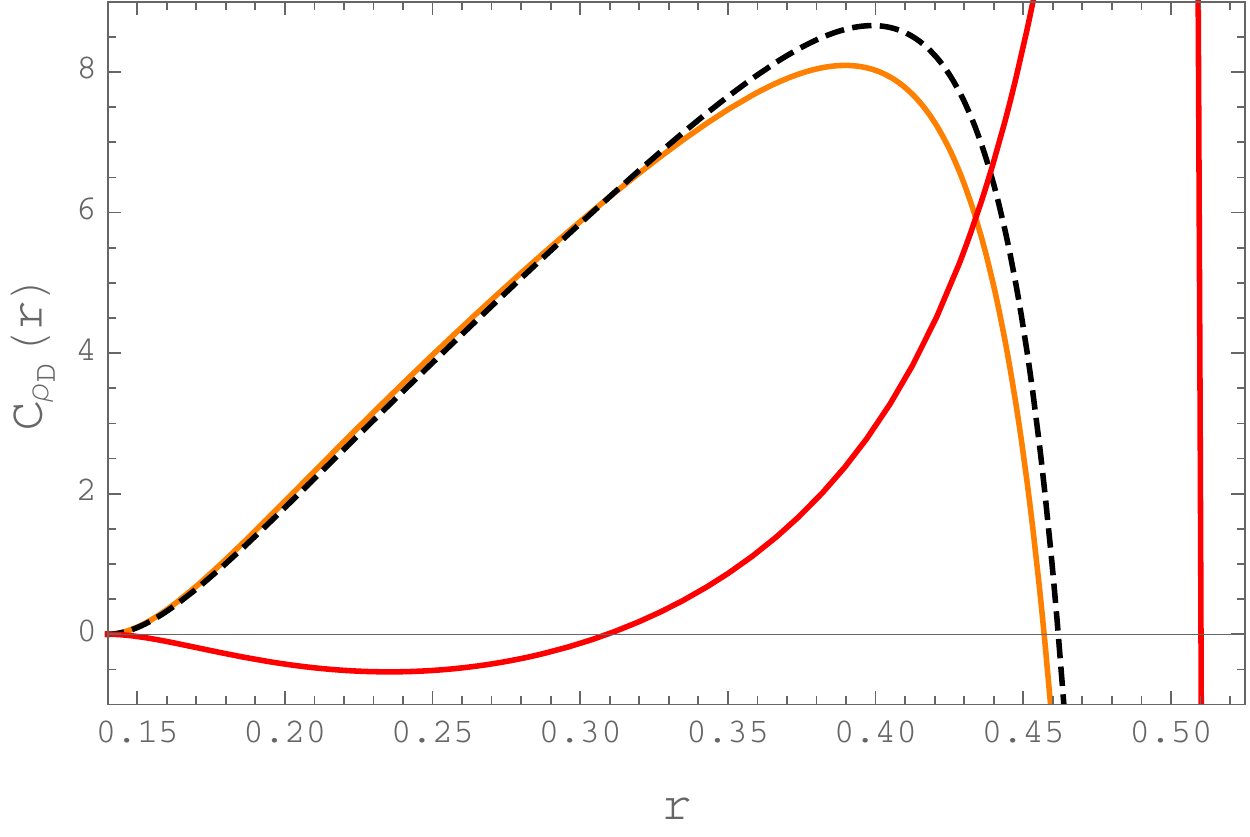}
\label{fig:subfigure2}}
\quad
\caption{Coefficients of the differential rate in the  $B\rightarrow X_c \tau \bar{\nu}_\tau$ decay as a function of the leptonic pair invariant mass $r$. 
The orange continuous and the black dashed lines stand for coefficients with LO and NLO precision, respectively. The continuous red line is the purely 
NLO contribution divided by $\alpha_s$.}
\label{fig:coefdGdq2bctv}
\end{figure}

\begin{figure}[ht]
\centering
\subfigure[Leading power coefficient.]{\includegraphics[scale=0.65]{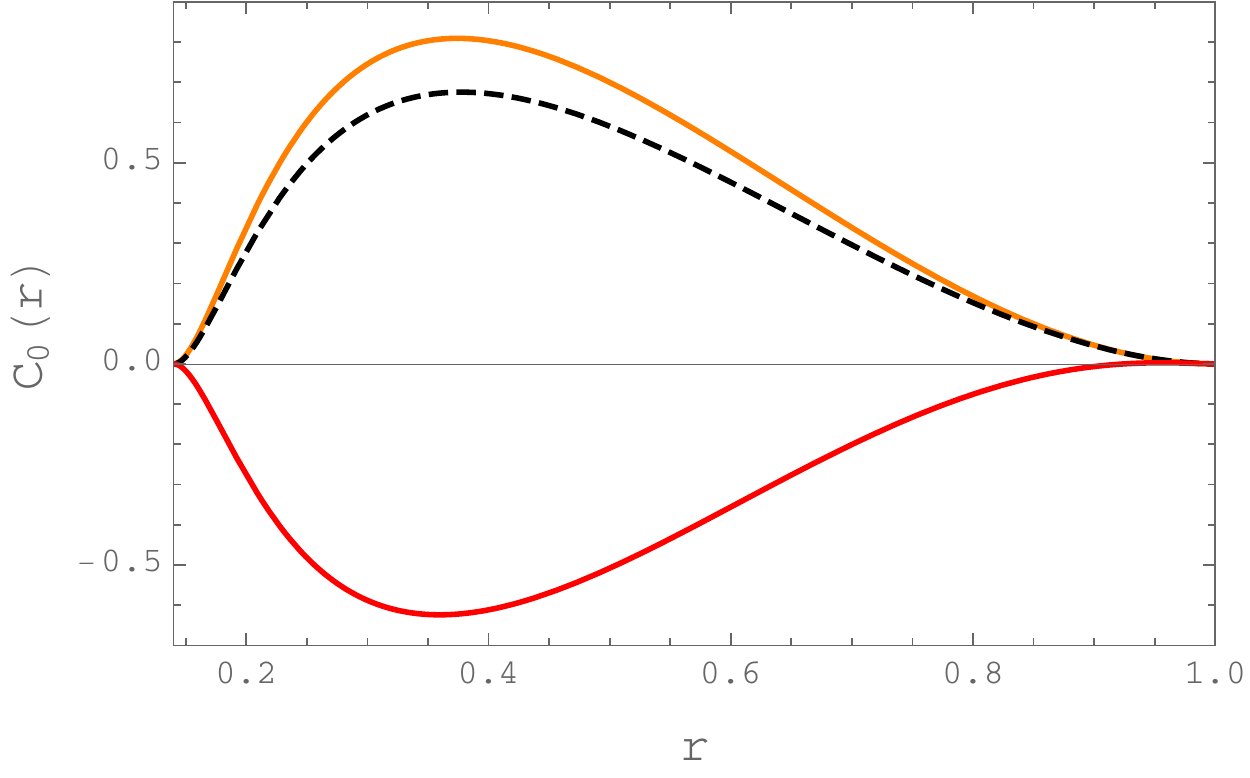}
\label{fig:subfigure1}}
\quad
\subfigure[Chromomagnetic operator coefficient.]{%
\includegraphics[scale=0.635]{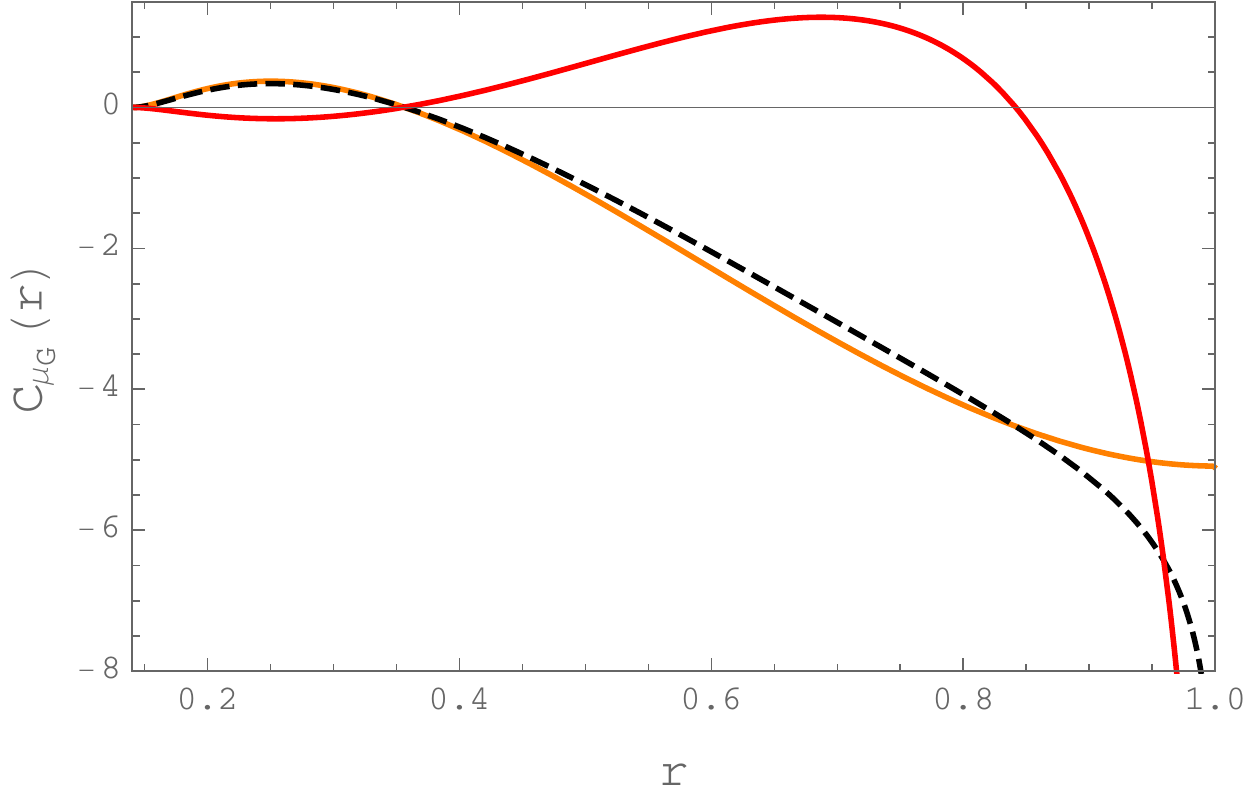}
\label{fig:subfigure3}}
\caption{Coefficients of the differential rate in the $B\rightarrow X_u \tau \bar{\nu}_\tau$ decay as a function of the leptonic pair invariant mass $r$. 
The orange continuous and the black dashed lines stand for coefficients with LO and NLO precision, respectively. The continuous red line is the purely 
NLO contribution divided by $\alpha_s$.}
\label{fig:coefdGdq2butv}
\end{figure}

In order to compare the coefficients of the differential rate with LO and NLO precision, as well as the shapes of the LO and NLO 
contributions, their dependence on the dilpeton invariant mass is shown in Fig.~\ref{fig:coefdGdq2bctv} for the 
$B\rightarrow X_{c} \tau \bar{\nu}_\tau$ decay channel, and in Fig.~\ref{fig:coefdGdq2butv} for the $B\rightarrow X_{u} \tau \bar{\nu}_\tau$ decay channel. 
Note that for the leading power coefficient the shapes of the LO and NLO contributions are very similar but with 
opposite sign. This is no longer true for the coefficients of the power corrections. In general, the $\alpha_s$ corrections to 
the coefficients of the power corrections are smaller at low $q^2$ than at high $q^2$.

\begin{figure}[ht]
\centering
\subfigure[Plot for $B \rightarrow X_c \tau \bar{\nu}_\tau$ decay.]{\includegraphics[scale=0.65]{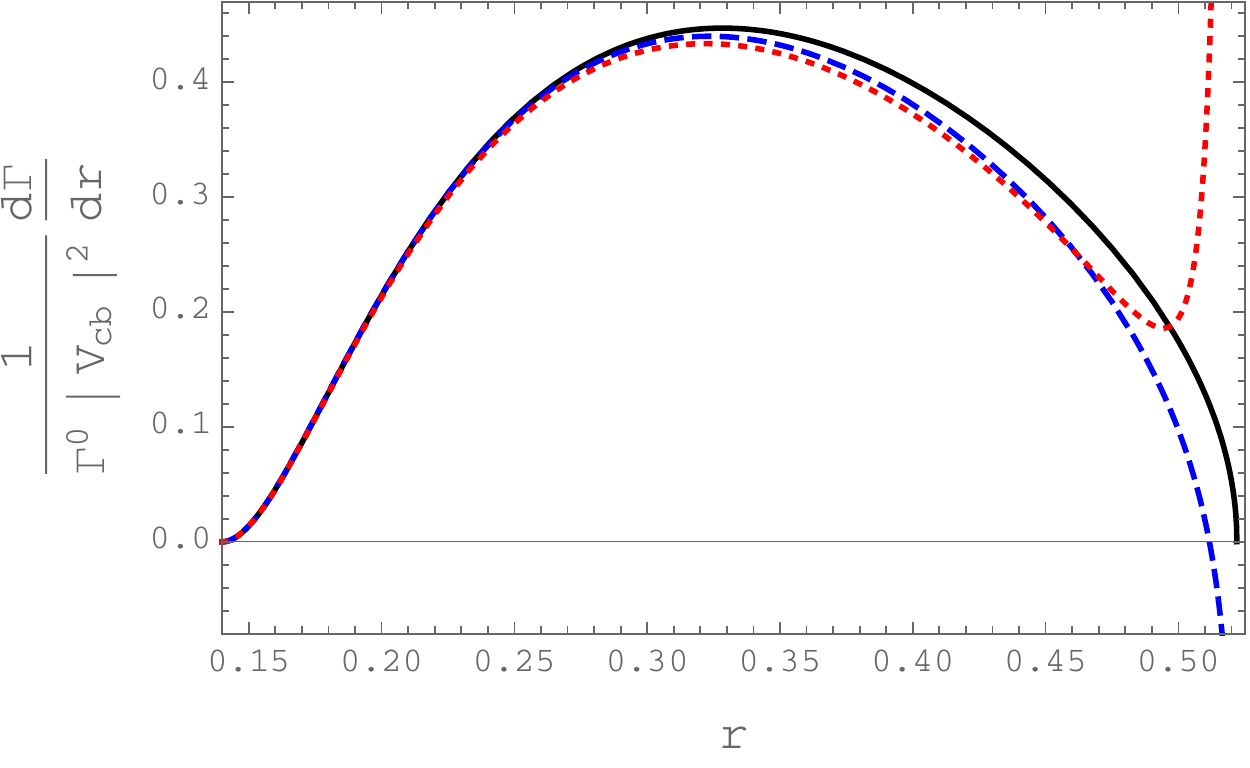}
\label{fig:subfigure1}}
\quad
\subfigure[Plot for $B \rightarrow X_u \tau \bar{\nu}_\tau$ decay.]{%
\includegraphics[scale=0.65]{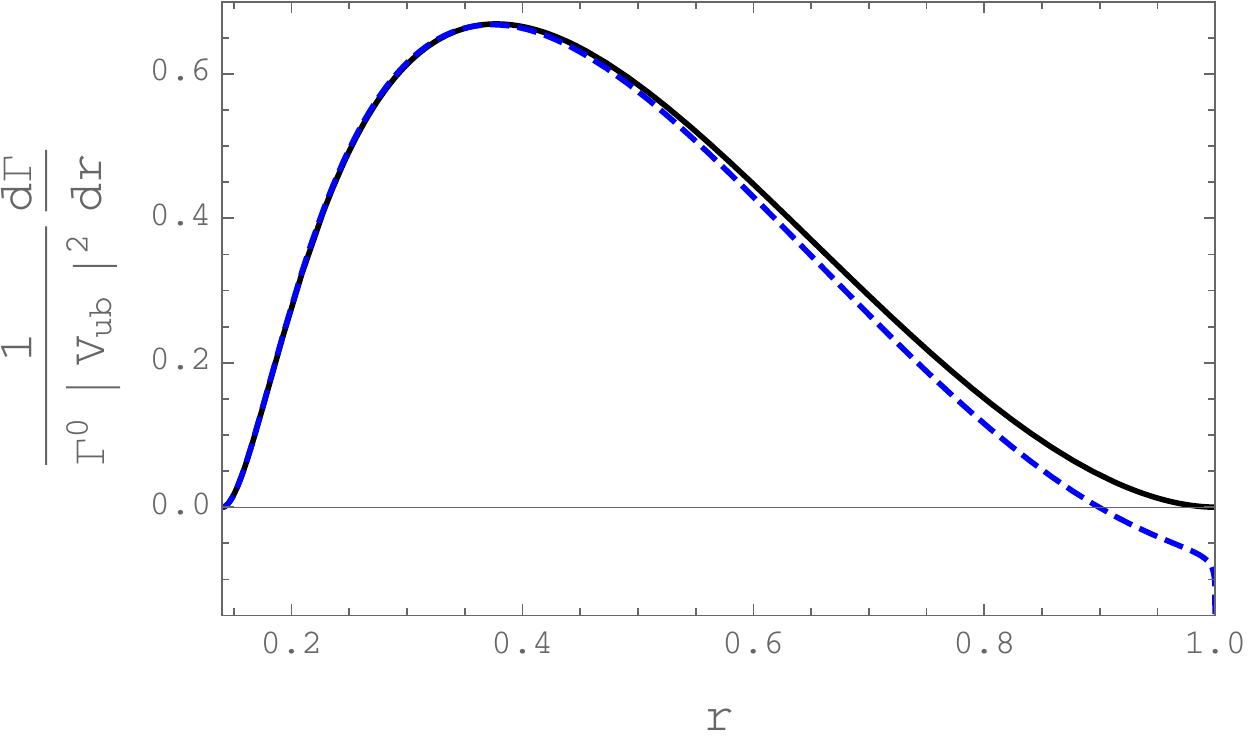}
\label{fig:subfigure3}}
\subfigure[Plot for $B \rightarrow X_c e \bar{\nu}_e$ decay.]{%
\includegraphics[scale=0.65]{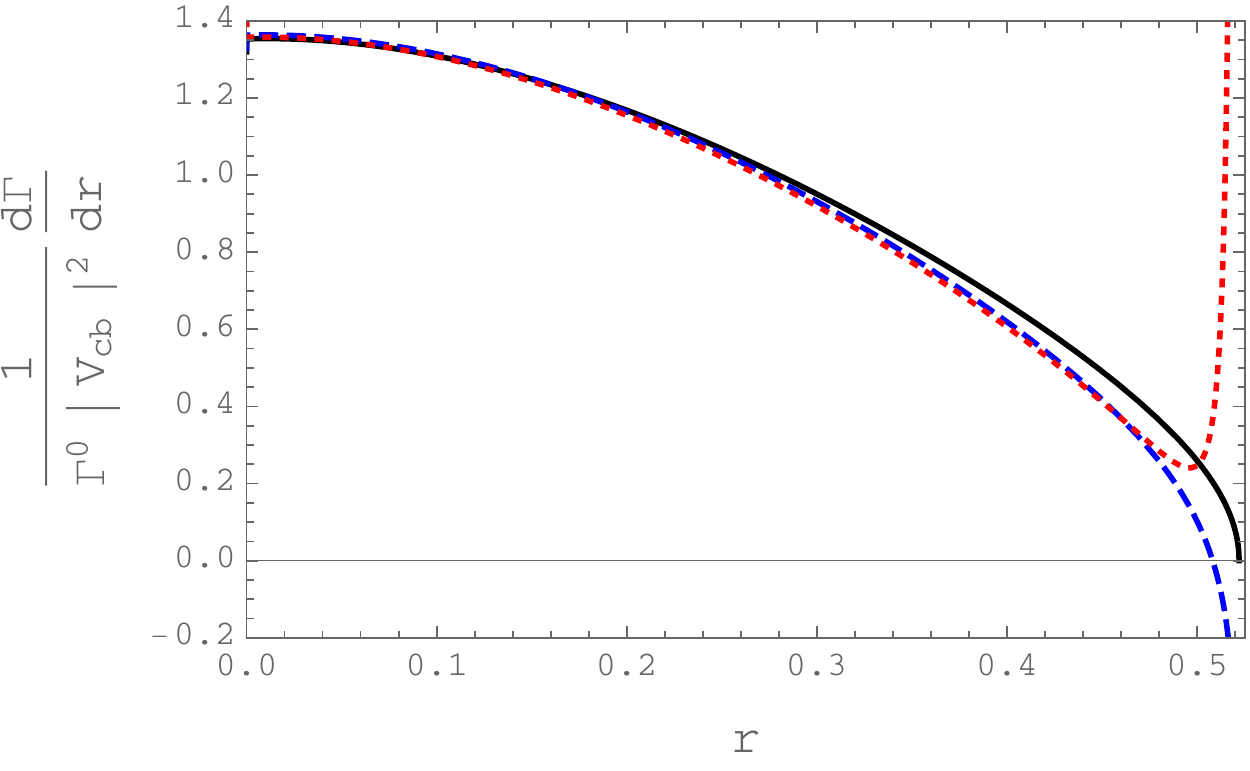}
\label{fig:subfigure2}}
\quad
\subfigure[Plot for $B \rightarrow X_u e \bar{\nu}_e$ decay.]{%
\includegraphics[scale=0.65]{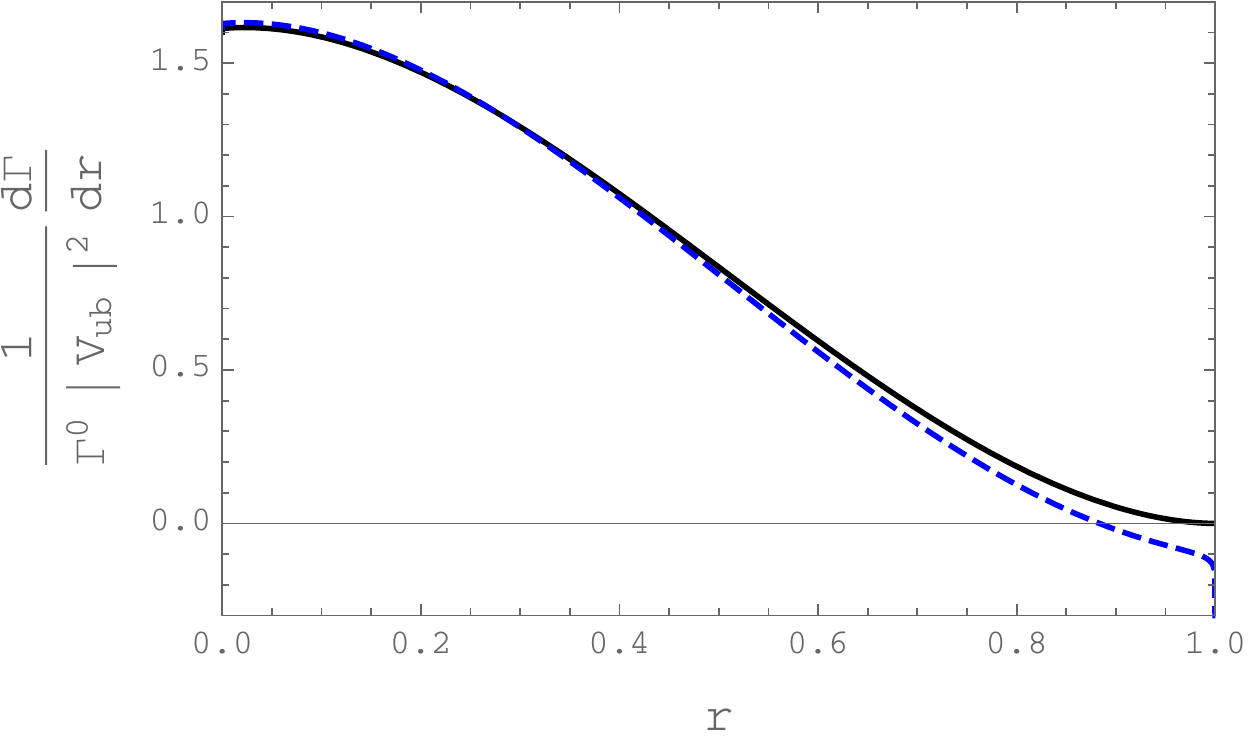}
\label{fig:subfigure4}}
\caption{Dependence on the lepton pair invariant mass $r$ of the differential rate for the $B \rightarrow X_q \ell \bar{\nu}_\ell$ decay channels 
by including subsequent power corrections with coefficients taken at NLO. The black continuous line stands for the leading power contribution, the blue dashed 
line includes $1/m_b^2$ corrections and the red dotted line includes $1/m_b^3$ corrections.}
\label{fig:difwidth}
\end{figure}

In order to see the importance of power corrections to the spectrum, Fig.~[\ref{fig:difwidth}] shows its dependence on the dilepton invariant 
mass by including subsequent power corrections for both decay channels, $B\rightarrow X_c \tau \bar{\nu}_\tau$ and 
$B\rightarrow X_u \tau \bar{\nu}_\tau$. It is interesting to compare the spectrum of semitauonic decays with their massless lepton counterparts. 
For this reason, plots are also presented for the $B\rightarrow X_c e \bar{\nu}_e$ and $B\rightarrow X_u e \bar{\nu}_e$ decay channels. Note that the 
lepton mass shifts the position of the peak to the right hand side of the spectrum and it also suppresses it. The latter statement is expected 
from experimental observation of the rate. 
The spectrum at large $q^2$ for $b\rightarrow u$ falls softer than for $b\rightarrow c$. In general, power corrections  
become larger at high $q^2$ and the IR behaviour at 
the endpoint becomes sharper for higher power corrections, pointing out the breakdown of the power expansion at large $q^2$. 

Some illustrative values for the different contributions to the moments $\bar{M}_n$, normalized moments $\hat{M}_n$, and ratios of moments 
$R^{q\ell/q'\ell'}_{n}$ between different decay channels are also provided, where
\begin{eqnarray}
 \bar{M}_n = \frac{M_n(B \rightarrow X_q \ell \bar{\nu}_\ell)}{\Gamma^0 |V_{qb}|^2}\,,\quad\;
 \hat{M}_n = \frac{M_n(B \rightarrow X_q \ell \bar{\nu}_\ell)}{M_0(B \rightarrow X_q \ell \bar{\nu}_\ell)}\,,\quad\; 
 R^{q\ell/q'\ell'}_{n} = \frac{|V_{q'b}|^2}{|V_{qb}|^2} \frac{M_n(B \rightarrow X_q \ell \bar{\nu}_\ell)}{M_n(B \rightarrow X_{q'} \ell' \bar{\nu}_{\ell'})}\,.
\end{eqnarray}
Numerical values are presented for the four decay channels $B\rightarrow X_q \ell \bar{\nu}_\ell$ with 
$q=c,\,u$ and $\ell= e,\,\tau$. This information is complementary to the one found in plots and helps to understand better the size of the 
corrections that have been computed, since they are of a few percent, and their size is difficult to illustrate in plots. Moreover, 
normalized moments and ratios might be useful as they are ideal observables to be compared to experiment because they do not depend on the absolute 
normalization of the rate which is very sensitive to the heavy quark mass (proportional to $m_b^5$). 

The results are given in the form (for normalized moments and ratios this is possible after re-expanding in $1/m_b$)
\begin{eqnarray}
  A_n  &=& \underbrace{A_{n,0}\bigg(1 - \frac{\mu_\pi^2}{2m_b^2}\bigg)}_{\rm part.}
 + \underbrace{A_{n,\mu_G} \bigg(\frac{\mu_G^2}{2m_b^2} - \frac{\rho_{LS}^3}{2m_b^3}\bigg)}_{\mu_G^2}
 \underbrace{- A_{n,\rho_D} \frac{\rho_D^3}{2m_b^3}}_{\rho_D^3} \,,
 \label{momdef}
\end{eqnarray}
where

\begin{eqnarray}   
 A_{n,i} &=& A_{n,i}^{\mbox{\scriptsize LO}} 
 + \frac{\alpha_s}{\pi} A_{n,i}^{\mbox{\scriptsize NLO}}\,,
 \label{spliALONLO}
\end{eqnarray}
and $A= \bar{M}_n,\,\hat{M}_n,\,R^{q\ell/q'\ell'}_{n}$. Numerical values are provided in tables for the coefficients on the right hand side of 
Eq.~(\ref{spliALONLO}), for the contribution of every power correction split in LO and NLO contributions (term by term sum), 
and for $A_n$ itself (total). 
Numerical values are presented in tables \ref{tab:mombctv}, \ref{tab:mombutv}, \ref{tab:mombcev}, and \ref{tab:mombuev} for the moments, 
in tables \ref{tab:Nmombctv}, \ref{tab:Nmombutv}, \ref{tab:Nmombcev}, and \ref{tab:Nmombuev} for the normalized moments, and in 
tables \ref{tab:ratcect}, \ref{tab:ratueut}, \ref{tab:ratutct}, and \ref{tab:ratuece} for the ratios.

As for moments, observe that in the $B\rightarrow X_c \tau \bar{\nu}_\tau$ decay the LO $1/m_b^3$ corrections represent a $8$-$24\%$ correction 
to the leading term. The $\alpha_s$ corrections to the Chromomagnetic and Darwin terms represent a $0.5$-$1.3\%$ correction to the leading term. 
In general, the corrections due to the Chromomagnetic and Darwin terms are similar in size. 
In the $B\rightarrow X_u \tau \bar{\nu}_\tau$ decay the $\alpha_s$ corrections to the Chromomagnetic term represent a 
$0.2$-$2.8\%$ correction to the leading term.

As for normalized moments, observe that in the $B\rightarrow X_c \tau \bar{\nu}_\tau$ decay the LO $1/m_b^3$ corrections represent 
a $4$-$16\%$ correction to the leading term. The $\alpha_s$ corrections to the Chromomagnetic and Darwin terms represent a 
$0.1$-$0.4\%$ and $0.6$-$2.6\%$ correction to the leading term, respectively. Note that, due to some numerical 
cancellations, the $\alpha_s/m_b^3$ corrections are comparable or even larger than the $\alpha_s$ corrections to the partonic term. 
In semitauonic decays the Darwin term corrections happen to be numerically larger than the Chromomagnetic term corrections, 
whereas in the case of massless leptons they are of similar size. 
In the $B\rightarrow X_u \tau \bar{\nu}_\tau$ decay the $\alpha_s$ corrections to the Chromomagnetic term represent a 
$0.6$-$3.5\%$ correction to the leading term.
 
As for ratios, observe that in $R_n^{ce/c\tau}$ the LO $1/m_b^3$ corrections represent 
a $5$-$6\%$ correction to the leading term. The $\alpha_s$ corrections to the Chromomagnetic and Darwin terms represent a 
$0.2$-$0.5\%$ and $0.5$-$0.6\%$ correction to the leading term, respectively. 
The Darwin term corrections happen to be numerically larger than the Chromomagnetic term corrections.
For the other ratios, only corrections up to $\alpha_s/m_b^2$ are available. The $\alpha_s$ corrections to the 
Chromomagnetic term represent a $0.1$-$3.6\%$ correction to the leading term. 
Note that in the ratios $R_n^{u\tau/c\tau}$ and $R_n^{ue/ce}$ numerical cancellations happen for the coefficient of the chromomagnetic 
term at LO which makes, in some cases, its NLO contribution to be numerically more important.
Finally, note that normalized moments and ratios are insensitive to $\mu_\pi^2$.

\clearpage
\newpage

\section*{Conclusions}
The current experimental value for $R(D^{(*)})$ in $B$ decays mediated by the semitauonic $b\rightarrow c \tau \bar{\nu}_\tau$ transition 
shows a more than $3\sigma$ deviation from the standard model prediction, which might point out the presence of new physics coupled 
to the $\tau$ lepton. In order to reveal the presence of new physics, the study of semitauonic decays is necessary by any means. 
In particular, inclusive decays may provide valuable 
complementary information to the one of exclusive decays, which motivates its investigation~\cite{Ligeti:2014kia,Ligeti:2021six}. 
If new physics is present in semitauonic decays, their study at the precision level might be necessary for its future establishment. 
Currently, the inclusive semitauonic decays have been poorly measured as they are experimentally challenging. 
However, their precise measurement should be possible in the near future by Belle II.
 
In this paper, analytical expressions for the dilepton invariant mass spectrum and total rate of inclusive semitauonic 
$B \rightarrow X_c \tau \bar{\nu}_\tau$ and $B \rightarrow X_u \tau \bar{\nu}_\tau$ decays have been obtained up to 
order $\alpha_s/m_b^3$ and $\alpha_s/m_b^2$, respectively. The dependence on the final state quark and the tau lepton masses is taken into account 
in its full analytical form. Analytical results are provided in the ancillary Mathematica notebook ``cobqtv.nb''. 
Therefore, the largest unknown contributions are the NNNLO corrections to the leading power term and the LO corrections to $1/m_b^4$ terms. 
For the case of massless leptons, these corrections have been considered for the total rate and a number of kinematical 
moments~\cite{Bigi:2009ym,Mannel:2010wj,Fael:2020tow,Fael:2022frj}.

It has been observed that for $q^2$ moments, normalized moments and ratios the LO $1/m_b^3$ corrections represent 
roughly a $10\%$ correction to the leading term. The $\alpha_s$ corrections to the Chromomagnetic and Darwin terms represent roughly a 
$1\%$ correction to the leading term. 

Precise theoretical predictions for the decay width, normalized moments and ratios can be derived by using our results. 
This requires nevertheless the use of some short distance mass like 
the 1S mass or the kinetic mass. The determination of the final estimates for the theory predictions are left to future publications.
It might be interesting to compare the theoretical predictions for the observables studied in this paper to future experimental measurements 
and look for possible discrepancies.

The results of this paper can be a used as a partial check and a breakthrough for future analytic calculations of NLO corrections 
to power suppressed terms in inclusive non-leptonic decays, in particular for the $b\rightarrow c \bar c s$ channel, which contains 
two massive particles in the final state, very much like in $b\rightarrow c\tau \bar{\nu}_\tau$.

\subsection*{Acknowledgments}
I thank Th.~Mannel, A.~A.~Pivovarov and Aleksey V.~Rusov for comments and discussions.
This research was supported by the Deutsche Forschungsgemeinschaft 
(DFG, German Research Foundation) under grant  396021762 - TRR 257 
``Particle Physics Phenomenology after the Higgs Discovery''.

\appendix

\section{Tables}
\label{App:tables}

\clearpage
\newpage

%MOMENTS
\begin{table}
\begin{center}
\begin{tabular}{ |c |c |c |c |c |c |c |c |c |}
 \hline
 {\footnotesize$\bar{M}_n$} & {\footnotesize$\bar{M}_{n,0}^{{\rm LO}}$} & {\footnotesize$\bar{M}_{n,0}^{{\rm NLO}}$} & {\footnotesize$\bar{M}_{n,\mu_G}^{{\rm LO}}$} & {\footnotesize$\bar{M}_{n,\mu_G}^{{\rm NLO}}$} & {\footnotesize$\bar{M}_{n,\rho_D}^{{\rm LO}}$} & {\footnotesize$\bar{M}_{n,\rho_D}^{{\rm NLO}}$} & {\footnotesize Term by term sum} & {\footnotesize Total}\\ 
 \hline
 & & & & & & & {\footnotesize$(0.1304 - 0.0127)_{\rm part.}$} &  \\ 
 {\footnotesize$n=0$}  &{\footnotesize$0.1316$} & {\footnotesize$-0.188$} & {\footnotesize$-1.09$} & {\footnotesize$1.4$} & {\footnotesize$10.9$} & {\footnotesize$9.4$} & {\footnotesize$ - ( 0.0094 - 0.0008)_{\mu_G^2}$} & {\footnotesize$0.0980$}\\ 
   & & & & & & & {\footnotesize$- ( 0.0105 + 0.0006)_{\rho_D^3}$} &  \\ 
 \hline
 & & & & & & & {\footnotesize$(0.0443 - 0.0042)_{\rm part.}$} &  \\ 
 {\footnotesize$n=1$}  & {\footnotesize$0.0447$} & {\footnotesize$-0.061$} & {\footnotesize$-0.51$} & {\footnotesize$0.58$} & {\footnotesize$5.6$} & {\footnotesize$5.0$} & {\footnotesize$ - (0.0044 - 0.0003)_{\mu_G^2}$} & {\footnotesize$0.0303$} \\ 
   & & & & & & & {\footnotesize$- ( 0.0054 + 0.0003)_{\rho_D^3}$} &  \\ 
 \hline
 & & & & & & & {\footnotesize$(0.0160 - 0.0015)_{\rm part.}$} &  \\ 
 {\footnotesize$n=2$}  & {\footnotesize$0.0162$} & {\footnotesize$-0.022$} & {\footnotesize$-0.24$} & {\footnotesize$0.25$} & {\footnotesize$2.9$} & {\footnotesize$2.6$} & {\footnotesize$ - (0.0021 - 0.0001)_{\mu_G^2}$} & {\footnotesize$0.0095$}\\ 
   & & & & & & & {\footnotesize$- (0.0028 + 0.0002)_{\rho_D^3}$} &  \\ 
 \hline 
 & & & & & & & {\footnotesize$(0.00612 - 0.00054)_{\rm part.}$} &  \\ 
 {\footnotesize$n=3$}  & {\footnotesize$0.00618$} & {\footnotesize$-0.008$} & {\footnotesize$-0.115$} & {\footnotesize$0.11$} & {\footnotesize$1.54$} & {\footnotesize$1.4$} & {\footnotesize$ - ( 0.00099 - 0.00007)_{\mu_G^2}$} & {\footnotesize$0.00308$} \\ 
   & & & & & & & {\footnotesize$- ( 0.00149 + 0.00009)_{\rho_D^3}$} &  \\ 
 \hline
\end{tabular}
\end{center}
\caption{Numerical values for the coefficients of moments in the $B\rightarrow X_c \tau \bar{\nu}_\tau$ channel.}
\label{tab:mombctv}
\end{table}
\begin{table}
\begin{center}
\begin{tabular}{ |c |c |c |c |c |c |c |}
 \hline
 {\footnotesize$\bar{M}_n$} & {\footnotesize$\bar{M}_{n,0}^{{\rm LO}}$} & {\footnotesize$\bar{M}_{n,0}^{{\rm NLO}}$} & {\footnotesize$\bar{M}_{n,\mu_G}^{{\rm LO}}$} & {\footnotesize$\bar{M}_{n,\mu_G}^{{\rm NLO}}$} & {\footnotesize Term by term sum} & {\footnotesize Total}\\ 
 \hline
  & & & & & &  \\
  {\footnotesize$n=0$}  &{\footnotesize$0.3640$} & {\footnotesize$-0.816$} & {\footnotesize$-1.824$} & {\footnotesize$-1.33$} &   {\footnotesize$(0.3607 - 0.0553)_{\rm part.}$} & {\footnotesize$0.2902$} \\ 
  & & & & &  {\footnotesize$ - ( 0.0145 + 0.0007)_{\mu_G^2}$} &  \\
\hline
 & & & & & &   \\ 
 {\footnotesize$n=1$}  & {\footnotesize$0.1689$} & {\footnotesize$-0.360$} & {\footnotesize$-1.45$} & {\footnotesize$-1.6$} &  {\footnotesize$(0.1673 - 0.0244)_{\rm part.}$} & {\footnotesize$0.1305$} \\ 
   & & & & &  {\footnotesize$ - ( 0.0115 + 0.0009)_{\mu_G^2}$} &  \\ 
\hline
 & & & & & &   \\ 
 {\footnotesize$n=2$}  & {\footnotesize$0.0885$} & {\footnotesize$-0.178$} & {\footnotesize$-1.19$} & {\footnotesize$-1.7$} &  {\footnotesize$(0.0877 - 0.0121)_{\rm part.}$} & {\footnotesize $0.0653$} \\ 
   & & & & &  {\footnotesize$- (0.0094 + 0.0009)_{\mu_G^2}$} &  \\ 
 \hline
 & & & & & &   \\ 
 {\footnotesize$n=3$}  & {\footnotesize$0.051$} & {\footnotesize$-0.097$} & {\footnotesize$-1.0$} & {\footnotesize$-1.8$} &  {\footnotesize$(0.051 - 0.007)_{\rm part.}$} & {\footnotesize$0.035$}\\ 
   & & & & &  {\footnotesize$ - ( 0.008 + 0.001)_{\mu_G^2}$} &  \\ 
 \hline
\end{tabular}
\end{center}
\caption{Numerical values for the coefficients of moments in the $B\rightarrow X_u \tau \bar{\nu}_\tau$ channel.}
\label{tab:mombutv}
\end{table}
\begin{table}
\begin{center}
\begin{tabular}{ |c |c |c |c |c |c |c |c |c |}
 \hline
 {\footnotesize$\bar{M}_n$} & {\footnotesize$\bar{M}_{n,0}^{{\rm LO}}$} & {\footnotesize$\bar{M}_{n,0}^{{\rm NLO}}$} & {\footnotesize$\bar{M}_{n,\mu_G}^{{\rm LO}}$} & {\footnotesize$\bar{M}_{n,\mu_G}^{{\rm NLO}}$} & {\footnotesize$\bar{M}_{n,\rho_D}^{{\rm LO}}$} & {\footnotesize$\bar{M}_{n,\rho_D}^{{\rm NLO}}$} & {\footnotesize Term by term sum} & {\footnotesize Total}\\ 
 \hline
 & & & & & & & {\footnotesize$(0.5649 - 0.0664)_{\rm part.}$} &  \\ 
 {\footnotesize$n=0$}  & {\footnotesize$0.5700$} & {\footnotesize$-0.979$} & {\footnotesize$-2.333$} & {\footnotesize$2.3$} & {\footnotesize$17.8$} & {\footnotesize$12.3$} & {\footnotesize$- (0.0202 - 0.0014)_{\mu_G^2}$} & {\footnotesize$0.4618$} \\ 
   & & & & & & & {\footnotesize$- ( 0.0171 + 0.0008)_{\rho_D^3}$} &  \\ 
\hline
 & & & & & & & {\footnotesize $(0.1160 - 0.0127)_{\rm part.}$} &  \\ 
 {\footnotesize $n=1$}  & {\footnotesize $0.1170$} & {\footnotesize $-0.188$} & {\footnotesize$-1.084$} & {\footnotesize$0.8$} & {\footnotesize$8.3$} & {\footnotesize$7.2$} & {\footnotesize$ - ( 0.0094 - 0.0005)_{\mu_G^2}$} & {\footnotesize $0.0859$} \\ 
   & & & & & & & {\footnotesize $- ( 0.0080 + 0.0005)_{\rho_D^3}$} &  \\ 
\hline
 & & & & & & & {\footnotesize $(0.0339 - 0.0035)_{\rm part.}$} &  \\ 
 {\footnotesize $n=2$}  & {\footnotesize $0.0342$} & {\footnotesize$-0.052$} & {\footnotesize$-0.50$} & {\footnotesize$0.3$} & {\footnotesize$4.23$} & {\footnotesize$3.8$} & {\footnotesize$ - (0.0043 - 0.0002)_{\mu_G^2}$} & {\footnotesize $0.0219$}\\ 
   & & & & & & & {\footnotesize$- (0.0041 + 0.0003)_{\rho_D^3}$} &  \\ 
 \hline
 & & & & & & & {\footnotesize$(0.01156 - 0.00116)_{\rm part.}$} &  \\ 
 {\footnotesize$n=3$}  & {\footnotesize$0.01167$} & {\footnotesize$-0.0171$} & {\footnotesize$-0.237$} & {\footnotesize$0.13$} & {\footnotesize$2.22$} & {\footnotesize$2.0$} & {\footnotesize$ - (0.00205 - 0.00008)_{\mu_G^2}$} & {\footnotesize $0.00616$} \\ 
   & & & & & & & {\footnotesize $- (0.00214 + 0.00013)_{\rho_D^3}$} &  \\ 
 \hline
\end{tabular}
\end{center}
\caption{Numerical values for the coefficients of moments in the $B\rightarrow X_c e \bar{\nu}_e$ channel.}
\label{tab:mombcev}
\end{table}
\begin{table}
\begin{center}
\begin{tabular}{ |c |c |c |c |c |c |c |}
 \hline
 {\footnotesize$\bar{M}_n$} & {\footnotesize$\bar{M}_{n,0}^{{\rm LO}}$} & {\footnotesize$\bar{M}_{n,0}^{{\rm NLO}}$} & {\footnotesize$\bar{M}_{n,\mu_G}^{{\rm LO}}$} & {\footnotesize$\bar{M}_{n,\mu_G}^{{\rm NLO}}$} & {\footnotesize Term by term sum} & {\footnotesize Total}\\ 
 \hline
 & & & & & &  \\ 
 {\footnotesize$n=0$}  & {\footnotesize$1$} & {\footnotesize$-2.413$} & {\footnotesize$-3$} & {\footnotesize$-3.5$} & {\footnotesize$ (0.991 - 0.164)_{\rm part.}$} & {\footnotesize$0.801$} \\ 
   & & & & & {\footnotesize$- (0.024 + 0.002)_{\mu_G^2}$} &  \\ 
\hline
 & & & & & &   \\ 
 {\footnotesize$n=1$}  & {\footnotesize$0.3$} & {\footnotesize$-0.674$} & {\footnotesize$-2.5$} & {\footnotesize$-3.3$} & {\footnotesize$(0.297 - 0.046)_{\rm part.}$} & {\footnotesize$0.229$} \\ 
   & & & & &  {\footnotesize$- (0.020 + 0.002)_{\mu_G^2}$} &  \\ 
\hline
 & & & & & &   \\ 
 {\footnotesize$n=2$}  & {\footnotesize$0.133$} & {\footnotesize$-0.28$} & {\footnotesize$-2$} & {\footnotesize$-3.4$} & {\footnotesize$ (0.132 - 0.019)_{\rm part.}$} & {\footnotesize$0.095$} \\ 
   & & & & & {\footnotesize$-(0.016 + 0.002)_{\mu_G^2}$} &  \\ 
 \hline
 & & & & & &  \\ 
 {\footnotesize$n=3$}  & {\footnotesize$0.0714$} & {\footnotesize$-0.14$} & {\footnotesize$-1.6$} & {\footnotesize$-3.4$} & {\footnotesize$(0.071 - 0.010)_{\rm part.}$} & {\footnotesize$0.046$}\\ 
   & & & & & {\footnotesize$-(0.013 + 0.002)_{\mu_G^2}$} &  \\ 
 \hline
\end{tabular}
\end{center}
\caption{Numerical values for the coefficients of moments in the $B\rightarrow X_u e \bar{\nu}_e$ channel.}
\label{tab:mombuev}
\end{table}
%NORMALIZED MOMENTS
\begin{table}
\begin{center}
\begin{tabular}{ |c |c |c |c |c |c |c |c |c |}
 \hline
 {\footnotesize$\hat{M}_n$} & {\footnotesize$\hat{M}_{n,0}^{{\rm LO}}$} & {\footnotesize$\hat{M}_{n,0}^{{\rm NLO}}$} & {\footnotesize$\hat{M}_{n,\mu_G}^{{\rm LO}}$} & {\footnotesize$\hat{M}_{n,\mu_G}^{{\rm NLO}}$} & {\footnotesize$\hat{M}_{n,\rho_D}^{{\rm LO}}$} & {\footnotesize$\hat{M}_{n,\rho_D}^{{\rm NLO}}$} & {\footnotesize Term by term sum} & {\footnotesize Total}\\ 
\hline
 & & & & & & & {\footnotesize$(0.3399 + 0.0012)_{\rm part.}$} &  \\ 
 {\footnotesize$n=1$}  & {\footnotesize$0.3399$} & {\footnotesize$0.017$} & {\footnotesize$-1.03$} & {\footnotesize$-0.5$} & {\footnotesize$14.3$} & {\footnotesize$32$} & {\footnotesize$-(0.0089 + 0.0003)_{\mu_G^2}$} & {\footnotesize$0.3160$} \\ 
   & & & & & & & {\footnotesize$-(0.0138 + 0.0021)_{\rho_D^3}$} &  \\ 
\hline
 & & & & & & & {\footnotesize$(0.1231 + 0.0008)_{\rm part.}$} &  \\ 
 {\footnotesize$n=2$}  & {\footnotesize$0.1231$} & {\footnotesize$0.012$} & {\footnotesize$-0.8$} & {\footnotesize$-0.4$} & {\footnotesize$12$} & {\footnotesize$27$} & {\footnotesize$-(0.0069 + 0.0002)_{\mu_G^2}$} & {\footnotesize$0.1034$} \\ 
   & & & & & & & {\footnotesize$- (0.0116 + 0.0018)_{\rho_D^3}$} &  \\ 
 \hline
 & & & & & & & {\footnotesize$(0.0470 + 0.0005)_{\rm part.}$} &  \\ 
 {\footnotesize$n=3$}  & {\footnotesize$0.0470$} & {\footnotesize$0.007$} & {\footnotesize$-0.483$} & {\footnotesize$-0.3$} & {\footnotesize$7.84$} & {\footnotesize$18$} & {\footnotesize$ - (0.0042 + 0.0002)_{\mu_G^2}$} & {\footnotesize$0.0343$} \\ 
   & & & & & & & {\footnotesize$-(0.0076 + 0.0012)_{\rho_D^3}$} &  \\ 
 \hline
\end{tabular}
\end{center}
\caption{Numerical values for the coefficients of the normalized moments in the $B\rightarrow X_c \tau \bar{\nu}_\tau$ channel.}
\label{tab:Nmombctv}
\end{table}
\begin{table}
\begin{center}
\begin{tabular}{ |c |c |c |c |c |c |c |}
 \hline
 {\footnotesize$\hat{M}_n$} & {\footnotesize$\hat{M}_{n,0}^{{\rm LO}}$} & {\footnotesize$\hat{M}_{n,0}^{{\rm NLO}}$} & {\footnotesize$\hat{M}_{n,\mu_G}^{{\rm LO}}$} & {\footnotesize$\hat{M}_{n,\mu_G}^{{\rm NLO}}$} & {\footnotesize Term by term sum} & {\footnotesize Total}\\ 
\hline
 & & & & & & \\ 
 {\footnotesize $n=1$}  & {\footnotesize$0.464$} & {\footnotesize$0.0508$} & {\footnotesize$-1.7$} & {\footnotesize$-6$} &  {\footnotesize$(0.464 + 0.003)_{\rm part.}$} & {\footnotesize $0.451$} \\ 
   & & & & &  {\footnotesize$- ( 0.013 + 0.003)_{\mu_G^2}$} &  \\ 
\hline
 & & & & & &   \\ 
 {\footnotesize$n=2$}  & {\footnotesize$0.243$} & {\footnotesize$0.06$} & {\footnotesize$-2$} & {\footnotesize$-8$} & {\footnotesize$(0.243 + 0.004)_{\rm part.}$} & {\footnotesize$0.227$} \\ 
   & & & & &  {\footnotesize$- ( 0.016 + 0.004)_{\mu_G^2}$} &  \\ 
 \hline
 & & & & & &   \\ 
 {\footnotesize$n=3$}  & {\footnotesize $0.141$} & {\footnotesize$0.05$} & {\footnotesize$-2$} & {\footnotesize$-9$} & {\footnotesize$(0.141 + 0.003)_{\rm part.}$} & {\footnotesize $0.123$} \\ 
   & & & & & {\footnotesize$-( 0.016 + 0.005)_{\mu_G^2}$} &  \\ 
 \hline
\end{tabular}
\end{center}
\caption{Numerical values for the coefficients of the normalized moments in the $B\rightarrow X_u \tau \bar{\nu}_\tau$ channel.}
\label{tab:Nmombutv}
\end{table}
\begin{table}
\begin{center}
\begin{tabular}{ |c |c |c |c |c |c |c |c |c |}
 \hline
 {\footnotesize $\hat{M}_n$} & {\footnotesize$\hat{M}_{n,0}^{{\rm LO}}$} & {\footnotesize$\hat{M}_{n,0}^{{\rm NLO}}$} & {\footnotesize$\hat{M}_{n,\mu_G}^{{\rm LO}}$} & {\footnotesize$\hat{M}_{n,\mu_G}^{{\rm NLO}}$} & {\footnotesize$\hat{M}_{n,\rho_D}^{{\rm LO}}$} & {\footnotesize$\hat{M}_{n,\rho_D}^{{\rm NLO}}$} & {\footnotesize Term by term sum} & {\footnotesize Total} \\ 
\hline
 & & & & & & & {\footnotesize $(0.2053 + 0.0016)_{\rm part.}$} &  \\ 
 {\footnotesize$n=1$}  & {\footnotesize$0.2053$} & {\footnotesize$0.023$} & {\footnotesize$-1.06$} & {\footnotesize$-1.1$} & {\footnotesize$8.2$} & {\footnotesize$21.6$} & {\footnotesize$ - ( 0.0092 + 0.0007)_{\mu_G^2}$} & {\footnotesize $0.1877$} \\ 
   & & & & & & & {\footnotesize$-(0.0079 + 0.0014)_{\rho_D^3}$} &  \\ 
\hline
 & & & & & & & {\footnotesize $(0.0599 + 0.0008)_{\rm part.}$} &  \\ 
 {\footnotesize$n=2$}  & {\footnotesize$0.0599$} & {\footnotesize$0.012$} & {\footnotesize$-0.64$} & {\footnotesize$-0.7$} & {\footnotesize$5.6$} & {\footnotesize$15$} & {\footnotesize$-(0.0055 + 0.0004)_{\mu_G^2}$} & {\footnotesize$0.0484$} \\ 
   & & & & & & & {\footnotesize$- (0.0054 + 0.0010)_{\rho_D^3}$} &  \\ 
 \hline
 & & & & & & & {\footnotesize$(0.0205 + 0.0004)_{\rm part.}$} &  \\ 
 {\footnotesize$n=3$}  & {\footnotesize$0.0205$} & {\footnotesize$0.0052$} & {\footnotesize$-0.33$} & {\footnotesize$-0.4$} & {\footnotesize$3.26$} & {\footnotesize$9$} & {\footnotesize$ - (0.0029 + 0.0002)_{\mu_G^2}$} & {\footnotesize$0.0141$} \\ 
   & & & & & & & {\footnotesize$- ( 0.0031 + 0.0006)_{\rho_D^3}$} &  \\ 
 \hline
\end{tabular}
\end{center}
\caption{Numerical values for the coefficients of the normalized moments in the $B\rightarrow X_c e \bar{\nu}_e$ channel.}
\label{tab:Nmombcev}
\end{table}
\begin{table}
\begin{center}
\begin{tabular}{ |c |c |c |c |c |c |c |}
 \hline
 {\footnotesize$\hat{M}_n$} & {\footnotesize$\hat{M}_{n,0}^{{\rm LO}}$} & {\footnotesize$\hat{M}_{n,0}^{{\rm NLO}}$} & {\footnotesize$\hat{M}_{n,\mu_G}^{{\rm LO}}$} & {\footnotesize$\hat{M}_{n,\mu_G}^{{\rm NLO}}$} & {\footnotesize Term by term sum} & {\footnotesize Total} \\ 
\hline
 & & & & & &   \\ 
 {\footnotesize$n=1$}  & {\footnotesize$0.3$} & {\footnotesize$0.05$} & {\footnotesize$-1.6$} & {\footnotesize$-6$} &  {\footnotesize$(0.300 + 0.003)_{\rm part.}$} & {\footnotesize$0.287$} \\ 
   & & & & &  {\footnotesize$-( 0.013 + 0.003)_{\mu_G^2}$} &  \\ 
\hline
 & & & & & &  \\ 
 {\footnotesize$n=2$}  & {\footnotesize$0.133$} & {\footnotesize$0.04$} & {\footnotesize$-1.6$} & {\footnotesize$-7$} &  {\footnotesize$(0.133 + 0.003)_{\rm part.}$} & {\footnotesize$0.119$} \\ 
   & & & & & {\footnotesize$- ( 0.013 + 0.004)_{\mu_G^2}$} &  \\ 
 \hline
 & & & & & &  \\ 
 {\footnotesize$n=3$}  & {\footnotesize$0.071$} & {\footnotesize$0.03$} & {\footnotesize$-1.4$} & {\footnotesize$-7$} & {\footnotesize$(0.071 + 0.002)_{\rm part.}$} & {\footnotesize$0.058$} \\ 
   & & & & &  {\footnotesize$-( 0.011 + 0.004)_{\mu_G^2}$} &  \\ 
 \hline
\end{tabular}
\end{center}
\caption{Numerical values for the coefficients of the normalized moments in the $B\rightarrow X_u e \bar{\nu}_e$ channel.}
\label{tab:Nmombuev}
\end{table}
%RATIOS
\begin{table}
\begin{center}
\begin{tabular}{ |c|c |c |c |c |c |c |c |c |}
 \hline
 {\footnotesize $R^{ce/c\tau}_{n}$} & {\footnotesize$R_{n,0}^{ce/c\tau,\,\mbox{\tiny LO}}$} & {\footnotesize$R_{n,0}^{ce/c\tau,\,\mbox{\tiny NLO}}$} & {\footnotesize$R_{n,\mu_G}^{ce/c\tau,\,\mbox{\tiny LO}}$} & {\footnotesize$R_{n,\mu_G}^{ce/c\tau,\,\mbox{\tiny NLO}}$} & {\footnotesize$R_{n,\rho_D}^{ce/c\tau,\,\mbox{\tiny LO}}$} & {\footnotesize$R_{n,\rho_D}^{ce/c\tau,\,\mbox{\tiny NLO}}$} & {\footnotesize Term by term sum} & {\footnotesize Total}\\ 
 \hline
 & & & & & & & {\footnotesize$(4.333 - 0.086)_{\rm part.}$} &  \\ 
 {\footnotesize$n=0$}  &{\footnotesize$4.333$} & {\footnotesize$-1.26$} & {\footnotesize$18$} & {\footnotesize$-12$} & {\footnotesize$-223.5$} & {\footnotesize$-432$} & {\footnotesize$ + (0.156 - 0.007)_{\mu_G^2}$} & {\footnotesize$4.639$}\\ 
   & & & & & & & {\footnotesize$+ (0.215 + 0.028)_{\rho_D^3}$} &  \\ 
\hline
 & & & & & & & {\footnotesize$(2.617 - 0.041)_{\rm part.}$} &  \\ 
 {\footnotesize$n=1$}  &{\footnotesize$2.617$} & {\footnotesize$-0.6$} & {\footnotesize$5.3$} & {\footnotesize$-15$} & {\footnotesize$-141$} & {\footnotesize$-247$} & {\footnotesize$+ (0.046 - 0.009)_{\mu_G^2}$} & {\footnotesize$2.765$}\\ 
   & & & & & & & {\footnotesize$+ (0.136 + 0.016)_{\rho_D^3}$} &  \\ 
\hline
 & & & & & & & {\footnotesize$(2.111 - 0.029)_{\rm part.}$} &  \\ 
 {\footnotesize$n=2$}  & {\footnotesize$2.111$} & {\footnotesize$-0.42$} & {\footnotesize$0.14$} & {\footnotesize$-19$} & {\footnotesize$-119$} & {\footnotesize$-185$} & {\footnotesize$ + (0.001 - 0.011)_{\mu_G^2}$} & {\footnotesize$2.199$}\\ 
   & & & & & & & {\footnotesize$+ (0.115 + 0.012)_{\rho_D^3}$} &  \\ 
 \hline
 & & & & & & & {\footnotesize$(1.89 - 0.02)_{\rm part.}$} &  \\ 
 {\footnotesize$n=3$}  &{\footnotesize$1.89$} & {\footnotesize$-0.3$} & {\footnotesize$-3$} & {\footnotesize$-24$} & {\footnotesize$-112$} & {\footnotesize$-155$} & {\footnotesize$ - (0.03 + 0.01)_{\mu_G^2}$} & {\footnotesize$1.950$}\\ 
   & & & & & & & {\footnotesize$+ (0.11 + 0.01)_{\rho_D^3}$} &  \\ 
 \hline
\end{tabular}
\end{center}
\caption{Numerical values for the coefficients of branching ratios $R^{ce/c\tau}_{n}$.}
\label{tab:ratcect}
\end{table}
\begin{table}
\begin{center}
\begin{tabular}{ |c |c |c |c |c |c |c |}
 \hline
 {\footnotesize$R^{ue/u\tau}_{n}$} & {\footnotesize$R_{n,0}^{ue/u\tau,\,\mbox{\tiny LO}}$} & {\footnotesize$R_{n,0}^{ue/u\tau,\,\mbox{\tiny NLO}}$} & {\footnotesize$R_{n,\mu_G}^{ue/u\tau,\,\mbox{\tiny LO}}$} & {\footnotesize$R_{n,\mu_G}^{ue/u\tau,\,\mbox{\tiny NLO}}$} & {\footnotesize Term by term sum} & {\footnotesize Total}\\ 
 \hline
 & & & & & &  \\ 
 {\footnotesize$n=0$}  &{\footnotesize$2.747$} & {\footnotesize$-0.47$} & {\footnotesize$5.5$} & {\footnotesize$10.2$} &  {\footnotesize$(2.747 - 0.032)_{\rm part.}$} & {\footnotesize$2.765$}\\ 
   & & & & &  {\footnotesize$+(0.044 + 0.006)_{\mu_G^2}$} &  \\ 
\hline
 & & & & & &   \\ 
 {\footnotesize$n=1$}  &{\footnotesize$1.777$} & {\footnotesize$-0.2$} & {\footnotesize$0.5$} & {\footnotesize$-4$} &  {\footnotesize$(1.777 - 0.014)_{\rm part.}$} & {\footnotesize$1.765$}\\ 
   & & & & &  {\footnotesize$+ (0.004 - 0.002)_{\mu_G^2}$} &  \\ 
\hline
 & & & & & &   \\ 
 {\footnotesize$n=2$}  &{\footnotesize$1.507$} & {\footnotesize$-0.138$} & {\footnotesize$-2.4$} & {\footnotesize$-15$}  & {\footnotesize$(1.507 - 0.009)_{\rm part.}$} & {\footnotesize$1.471$}\\ 
   & & & & &  {\footnotesize$- (0.019 + 0.008)_{\mu_G^2}$} &  \\ 
 \hline
 & & & & & &   \\ 
 {\footnotesize$n=3$}  &{\footnotesize$1.395$} & {\footnotesize$-0.11$} & {\footnotesize$-5$} & {\footnotesize$-29$} & {\footnotesize$(1.395 - 0.008)_{\rm part.}$} & {\footnotesize$1.331$}\\ 
   & & & & &  {\footnotesize$-(0.040 + 0.016)_{\mu_G^2}$} &  \\ 
 \hline
\end{tabular}
\end{center}
\caption{Numerical values for the coefficients of branching ratios $R^{ue/u\tau}_{n}$.}
\label{tab:ratueut}
\end{table}
\begin{table}
\begin{center}
\begin{tabular}{ |c |c |c |c |c |c |c |}
 \hline
 {\footnotesize$R^{u\tau/c\tau}_{n}$} & {\footnotesize$R_{n,0}^{u\tau/c\tau,\,\mbox{\tiny LO}}$} & {\footnotesize$R_{n,0}^{u\tau/c\tau,\,\mbox{\tiny NLO}}$} & {\footnotesize$R_{n,\mu_G}^{u\tau/c\tau,\,\mbox{\tiny LO}}$} & {\footnotesize$R_{n,\mu_G}^{u\tau/c\tau,\,\mbox{\tiny NLO}}$} & {\footnotesize Term by term sum} & {\footnotesize Total}\\ 
 \hline
 & & & & & &  \\ 
 {\footnotesize$n=0$}  &{\footnotesize$2.77$} & {\footnotesize$-2.25$} & {\footnotesize$9$} & {\footnotesize$-44$} & {\footnotesize$(2.77 - 0.15)_{\rm part.}$} & {\footnotesize$2.67$}\\ 
   & & & & &  {\footnotesize$+ ( 0.07 - 0.02)_{\mu_G^2} $} &  \\ 
\hline
 & & & & & &  \\ 
 {\footnotesize$n=1$}  &{\footnotesize$3.78$} & {\footnotesize$-3$} & {\footnotesize$10$} & {\footnotesize$-102$} &  {\footnotesize$(3.78 - 0.20)_{\rm part.}$} & {\footnotesize$3.60$}\\ 
   & & & & &  {\footnotesize$+ ( 0.08 - 0.06)_{\mu_G^2}$} &  \\ 
\hline
 & & & & & &  \\ 
 {\footnotesize$n=2$}  & {\footnotesize$5.47$} & {\footnotesize$-3.8$} & {\footnotesize$7$} & {\footnotesize$-237$} &  {\footnotesize$(5.47 - 0.26)_{\rm part.}$} & {\footnotesize$5.14$}\\ 
   & & & & &  {\footnotesize$+ (0.06 - 0.13)_{\mu_G^2}$} &  \\ 
 \hline
 & & & & & &   \\ 
 {\footnotesize$n=3$}  &{\footnotesize$8.29$} & {\footnotesize$-5.1$} & {\footnotesize$-6.8$} & {\footnotesize$-547$} &  {\footnotesize$(8.29 - 0.35)_{\rm part.}$} & {\footnotesize$7.59$}\\ 
   & & & & & {\footnotesize$-( 0.05 + 0.30)_{\mu_G^2}$} &  \\ 
 \hline
\end{tabular}
\end{center}
\caption{Numerical values for the coefficients of branching ratios $R^{u\tau/c\tau}_{n}$.}
\label{tab:ratutct}
\end{table}
\begin{table}
\begin{center}
\begin{tabular}{ |c |c |c |c |c |c |c |}
 \hline
 {\footnotesize$R^{ue/ce}_{n}$} & {\footnotesize$R_{n,0}^{ue/ce,\,\mbox{\tiny LO}}$} & {\footnotesize$R_{n,0}^{ue/ce,\,\mbox{\tiny NLO}}$} & {\footnotesize$R_{n,\mu_G}^{ue/ce,\,\mbox{\tiny LO}}$} & {\footnotesize$R_{n,\mu_G}^{ue/ce,\,\mbox{\tiny NLO}}$} & {\footnotesize Term by term sum} & {\footnotesize Total}\\ 
 \hline
 & & & & & &  \\ 
 {\footnotesize$n=0$}  & {\footnotesize$1.754$} & {\footnotesize$-1.220$} & {\footnotesize$1.9$} & {\footnotesize$-15$} & {\footnotesize$ (1.754 - 0.084)_{\rm part.}$} &{\footnotesize $1.677$}\\ 
   & & & & & {\footnotesize$+ (0.015 - 0.008)_{\mu_G^2}$} &  \\ 
\hline
 & & & & & &  \\ 
 {\footnotesize$n=1$}  &{\footnotesize$2.56$} & {\footnotesize$-1.6$} & {\footnotesize$2$} &{\footnotesize $-57$} & {\footnotesize$ (2.56 - 0.11)_{\rm part.}$} & {\footnotesize$2.44$}\\ 
   & & & & & {\footnotesize$+ ( 0.02 - 0.03)_{\mu_G^2}$} &  \\ 
\hline
 & & & & & &  \\ 
 {\footnotesize$n=2$}  &{\footnotesize$3.90$} & {\footnotesize$-2.3$} & {\footnotesize$-1.2$} & {\footnotesize$-170$} & {\footnotesize$ (3.90 - 0.16)_{\rm part.}$} & {\footnotesize$3.64$}\\ 
   & & & & & {\footnotesize$- (0.01 + 0.09)_{\mu_G^2}$} &  \\ 
 \hline
 & & & & & &  \\ 
 {\footnotesize$n=3$}  &{\footnotesize$6.1$} & {\footnotesize$-3$} & {\footnotesize$-17$} & {\footnotesize$-449$} & {\footnotesize$ (6.1 - 0.2)_{\rm part.}$} & {\footnotesize$5.6$}\\ 
   & & & & & {\footnotesize$- (0.1 + 0.2)_{\mu_G^2}$} &  \\ 
 \hline
\end{tabular}
\end{center}
\caption{Numerical values for the coefficients of branching ratios $R^{ue/ce}_{n}$.}
\label{tab:ratuece}
\end{table}

\end{document}